\documentclass[prb,twocolumn,floats,ap`s,showpacs,epsf,amsmath,amssymb]{revtex4}
\usepackage{epsfig}

\newcommand{\be}{\begin{equation}}
\newcommand{\ee}{\end{equation}}
\newcommand{\bea}{\begin{eqnarray}}
\newcommand{\eea}{\end{eqnarray}}

 \newcommand{\p}{\partial}
\newcommand{\s}{\sigma} 
 \newcommand{\la}{\langle}
\newcommand{\ra}{\rangle} \newcommand{\rd}{\mbox{d}}
\newcommand{\ri}{\mbox{i}} \newcommand{\re}{\mbox{e}}
 
 \newcommand{\up}{\uparrow}
\newcommand{\down}{\downarrow}

\begin{document}
%\draft
\title{Excitation spectrum  of doped two-leg ladders: A field theory
analysis}

\author{ D. Controzzi$^1$ and A.  M. Tsvelik$^2$} \affiliation{$^1$
International School for Advanced Studies and INFN, via Beirut 4,
34014 Trieste, Italy, \\ $^2$ Department of Physics, Brookhaven
National Laboratory, Upton, NY 11973-5000, USA}

\date{\today}

\begin{abstract}
We apply quantum field theory to study the excitation 
spectrum of doped two-leg ladders. It follows from our analysis  that
throughout most of the phase diagram  the spectrum consists of
degenerate quartets of kinks and anti-kinks and a multiplet of vector
particles split according to the symmetry of the problem as 3 + 2
+1. This basic picture experiences corrections when one moves through
the phase diagram. In some regions the
splitting may become very small and in others it is so large that
some multiplets are pushed in the continuum and become unstable. At
second  order transition lines masses of certain particles
vanish. Very close to the first order transition line additional
generations of particles emerge. Strong interactions in some sectors
may generate additional bound states (like breathers) in the asymmetric
charge sector. We briefly describe the properties of various
correlation functions in different phases.

\end{abstract}

\maketitle

\sloppy
\section{Introduction}

The problem  of  ladder-like materials  has attracted a lot of
attention since the 
original paper by Dagotto and Rice \cite{Dagotto}. Important experimental
realizations of two-leg ladder systems include, for example, the 
famous 'telephone number' compound Sr$_{14-x}$Ca$_x$Cu$_{24}$O$_{41}$ which 
has a spin gap and exhibits a transition from Charge Density Wave (CDW) to 
superconducting (SC) state under the increase of Ca concentration 
\cite{uehara,imai,eccleston,arpes,gorshunov,abba,abba2,girsh}. 
Besides the experimental relevance the study of this 
problem gives rise to  many questions  of rather general character.  
Even if ladder systems represent only the first step from a purely 
one-dimensional world into higher dimensions, this step introduces  a
lot of new 
interesting physics. In addition, the problem of doped spin ladders is
just a particular case of a more  general 
problem  of multi-orbital 
quasi-one-dimensional models \cite{1D}, and  
an increase in the number of orbitals dramatically  
increases the complexity of the lattice Hamiltonian,  rising legitimate 
questions about universality. 

In this paper we focus on the problem of doped two-leg
ladders. Assuming that the spectral gaps are much smaller than the
bandwidth (the applicability of these assumption to real systems is
discussed in the end of the paper), we study the low-energy physics
using the field theory approach.   
The  intensive theoretical research conducted on this subject in 
the last eight years has established the following facts.\\
{\em A. Strong-weak tunneling duality}. 
The form of the effective action describing the low energy behavior of the
system is independent 
(up to a simple operator transformation) on whether one takes into account the
inter-chain tunneling or considers just the inter-chain interactions. One
arrives to this conclusion comparing effective actions derived in the  limit
of strong tunneling  (see, for example, 
Ref.s~\onlinecite{BF96,Lin,Fradkin03,TS04}) with the 
theories derived in the limit of weak tunneling \cite{ShTs96,Azaria}.  
The strong-weak tunneling  duality may be  a boon for numerical calculations 
allowing to extract information about the excitation spectrum in one parameter 
range by performing  actual calculations in the other. \\
{\em B. Superconductivity-CDW duality}. The phase diagram includes the 
Tomonaga-Luttinger (TL) type phase as well as phases with spectral
gaps. The only stable TL phase is the one where all modes are
gapless. As soon as an attractive interaction appears in one channel it induces
attraction in all other ones generating spectral gaps for all modes
except the symmetric charge mode. The strong coupling  phases are 
characterized by power law
correlations for particular operators (order parameters). 
These  phases are classified either as SC
or CDW.  For the two chain model
there are two SC phases ($s$ and $d$) and two CDW phases (also $s$ and
$d$, the latter phase also being known as Orbital Antiferromagnet
\cite{OAF} or Staggered Flux \cite{affleck} phase). 
Such classification is valid only for a weak repulsion when the Luttinger
parameter $K_c$ characterizing the gapless charge mode, is close to
one.  At  $K_c < 1/2$ the Wigner crystal phase with $4k_F$ density
correlations competes with the SC phases
\cite{giamarchi.book,marston}. The low-energy
Hamiltonians in different sectors differ by the sign of certain
coupling  constants and transform  to each other by canonical
transformations  of the fields. These transformations  realize  
authomorphisms of the O(6) group \cite{Azaria}.\\
{\em C.  Emergent attractive interactions. }
 As follows both from  the  numerical \cite{Troyer96,Poilblanc}
 and analytic calculations  based on the renormalization group (RG)
 analysis \cite{Lin,Fradkin03,Azaria,Orignac97,Azaria2}, 
the system at low energies may enter into a 
strong coupling regime even if all bare couplings are repulsive. 
In that case the system passes  through an
intermediate weak coupling regime. If  $E^*$ is the  energy at which
 the weak  coupling  is reached, than it can be shown that the strong
 coupling  regime is achieved at energies of the order  of
 ${E^*}^2/\Lambda$,  where $\Lambda$ is the bare ultraviolet
 cut-off. It is probably safe to say that the emergent attraction
 leads to  small gaps. 

It was first suggested in Ref.~\onlinecite{Lin} that if the bare couplings
are not very large, the low energy sector of a two-leg ladder is
described by a universal Hamiltonian with a symmetry larger than the 
symmetry of the lattice model. This corresponds to the
O(6)$\times$U(1) symmetry in the doped case and the O(8) symmetry at
half filling. The original suggestion was based on the observation
that the RG  flows of different coupling constants converge in the
strong coupling limit to the same asymptotics corresponding 
to a higher symmetry. Convergence of RG flows  has also been  found to occur 
in other systems \cite{Azaria2}
and is frequently associated with  Dynamical Symmetry Enlargement (DSE). 
A careful discussion of DSE for some specific models can be found in 
Ref.~\onlinecite{kls}. This problem is addressed also here 
matching the RG analysis, valid at weak coupling, with methods valid 
%large $N$ and Form-factor Perturbation Theory 
at strong coupling. The conclusion is that in models where the number
of particle flavors is not large, there are conditions when one may
indeed expect a small  splitting of particle multiplets. Such
conditions exist in systems with weak backscattering, like  carbon
nanotubes.

The goal of this paper is to use a quantum field theory approach
to describe the phase diagram and excitation spectrum 
of a  doped two-leg Hubbard-type model.
We combine RG with methods
suitable to study the strong coupling limit, like $1/N$ expansion \cite{1/N},
exact solutions
and perturbation theory around specific integrable points in the
parameter space of the effective field theory \cite{dms}.
We also discuss the conditions of  validity of the field
theory description. As said above, this description is valid
when the correlation length is much larger then the ultra-violet (UV)
cut-off. This regime is known as the {\em scaling limit}.

The paper is organized as follows. In the next Section we introduce
the effective low-energy description of doped
two-leg ladders, the so-called  generalized $O(6)$ Gross-Neveu model, 
and briefly analyze  the RG equations.
In Sec.~\ref{sec:phases} we discuss the 
possible strong coupling phases of the model 
and in Sec.~\ref{sec:symmetries} we
describe its symmetries  and
identify the phase boundaries. The spectrum of generalized Gross-Neveu models
is discussed in Sec.~\ref{sec:semiclassical} using the large $N$ 
approximation. Some specific
points of  the phase diagram are described by integrable models with
the O(6) GN model being the most prominent. These integrable points
are identified in Sec.~\ref{sec:int.points}. In the vicinity of
these points we study the spectrum using a specific strong-coupling 
perturbation theory  described in Sec. ~\ref{sec:FFPT}. The
spectrum close to the phase boundaries is summarized in
Sec.~\ref{sec:spectrumatboundaries}.  
Sec. ~\ref{sec:o3timeso3} is devoted to the study of a simplified 
O(3)$\times$O(3)-symmetric model where the  RG equations can be
integrated  explicitly. In Sec.~\ref{sec:corrfunc} we briefly consider
the structure of some correlation functions. We summarize the main
results and discuss the experimental relevance in the last section.
The paper has several Appendixes.

\section{The model}

The model in its original formulation includes electron creation and
annihilation operators $c^\dagger_{i,l,\s},c_{i,l,\s}$ labeled by
chain indices, $l=1,2$, and spin indices, $\s=+1$ for spin up ($\up$) and 
$\s=-1$ for spin down ($\down$). For instance, the
extended Hubbard model has the standard form 
\bea H&=&-t_\parallel
\sum_{i,l,\s} \left ( c^\dagger_{i,l,\s} c_{i+1,l,\s}+h.c. \right)
+U\sum_{i,l} n_{i,l,\up} n_{i,l,\down}\nonumber \\ &&- t_\perp
\sum_{i,l,\s} \left ( c^\dagger_{i,1,\s} c_{i,2,\s}+h.c. \right)
\nonumber\\ &&+ V_\parallel \sum_{i,l} n_{i,l} n_{i+i,l} +V_\perp
\sum_{i,l} n_{i,1} n_{i,2},
\label{H.lattice}
\eea 
where, as usual, the parameters $t_\parallel$, $t_\perp$ are the
hopping matrix elements along and between the chains, $U$ is the
on-site repulsion and $V_\parallel$, $V_\perp$ the next-nearest neighbor 
interactions, with $n_{i,l,\s}=c^\dagger_{i,l,\s} c_{i,l,\s}$ and
$n_{i,l}=n_{i,l,\up}+n_{i,l,\down}$. The lattice Hamiltonian
(\ref{H.lattice}) has a $U(1)\times SU(2) \times \mathbb{Z}_2$
symmetry. If necessary, one can include  also exchange
interactions.

\subsection{Low energy field theory}

As we have mentioned in the introduction, the effective field theory,
describing the low energy behavior of the Hamiltonian
(\ref{H.lattice}), is largely independent on whether one considers
strong interactions and weak inter-chain tunneling, as it was done, for
example, in Ref.s~\onlinecite{ShTs96,Azaria}, or diagonalizes the inter-chain
hopping first and treats the bare interactions as weak, as it has been
done in the majority of other papers. To be precise, there are two
differences: (i) in one case the fields are labeled by chain indices
and in the other by transverse wave vectors $q = 0,\pi$, (ii) the
Hamiltonians of both sectors are related to each other by the
particle-hole transformation.
%corresponding to the automorphism of the
%O(6) group \cite{Azaria}.

As far as the analysis of the phase diagram and the spectrum are concerned, it
is more advantageous to have a low energy description in the Majorana
fermion representation.  The Majorana formulation is derived by
bosonization and subsequent refermionization of the original
Hamiltonian (the procedure was introduced in Ref.~\onlinecite{snt}, see also
\onlinecite{book}) and, for the specific model (\ref{H.lattice}), can be
found, for instance, in Ref.s~\onlinecite{TS04,giamarchi.book}. 
Away from half-filling, the
resulting low energy field theory consists of two parts: one contains
a decoupled symmetric charge mode $\Phi_c^{(+)}$ and the other 
contains all other fields. The latter can be written in terms of 6
Majorana fermions as
\begin{subequations}
\label{effectiveH}
\bea H = H_0 + V,
\label{mod} 
\eea where $H_0$ is the free part \bea H_0 = &-& \frac{\ri
}{2}\sum_{a=1}^3v_s(\chi^a_R\p_x\chi^a_R - \chi^a_L\p_x\chi^a_L)
\nonumber \\ &-& \frac{\ri}{2}\sum_{a=1}^3v_a(\xi^a_R\p_x\xi^a_R -
\xi^a_L\p_x\xi^a_L) \eea and $V$ describes the interaction between the
right and the left moving Majorana fermions $\chi^a_{R,L}$ and
$\xi^a_{R,L}$ \bea V &=& - g_{\s +}(\chi^a_R\chi^a_L)^2 -
g_{\rho-}\left[(\xi^1_R\xi^1_L) + (\xi^2_R\xi^2_L)\right]^2 \nonumber
\\ &&- 2(\chi^a_R\chi^a_L) \left\{g_{\s -}(\xi^3_R\xi^3_L) +
g_{c,st}\left[(\xi^1_R\xi^1_L) + (\xi^2_R\xi^2_L)\right]\right\}
\nonumber\\ &&- 2g_{c,ss}(\xi^3_R\xi^3_L)\left[(\xi^1_R\xi^1_L) +
(\xi^2_R\xi^2_L)\right],
\label{int}
\eea
\end{subequations}
(the summation over repeated indices is assumed).  The symmetry of the
continuum Hamiltonian is $U(1)\times U(1)\times SU(2)\times
\mathbb{Z}_2$, being somewhat higher than the symmetry of the lattice
model. As we have already mentioned, one $U(1)$ field (the total
charge mode $\Phi_c^{(+)}$ ) is decoupled and is not shown in the
above Hamiltonian, while the second $U(1)$ symmetry emerges only
asymptotically at lower energies. The mode $\Phi_c^{(+)}$ is responsible for 
 high conductivity along the ladders observed in Sr$_{14-x}$Ca$_{x}$Cu$_{24}$O$_{41}$ \cite{girsh}.  The SU(2) triplet $\chi^a$ and the
fermion $\xi^3$ are made of bosonic fields of the spin sector. They
reflect the $SU(2) \times \mathbb{Z}_2$ symmetry of the spin sector
and also appear in the low energy description of the two-leg
Heisenberg ladder \cite{snt}.  The situation here differs from the
Heisenberg ladder since no explicit mass term is present in the
Hamiltonian and the masses are generated dynamically.  The fermionic
doublet $\xi_{1,2}$ describes the asymmetric charge mode, which we
will denote $\Theta_c^{(-)}$.

The original observables are nonlocal in terms of the Majorana
fermions (vector particles). The latter particles, if remain stable,
represent collective excitations of the system. We emphasize that the
vector multiplet in (\ref{effectiveH}) 
is naturally split into sub-multiplets as 3 + 1 +2. The
model (\ref{effectiveH}) can also be represented as six critical Ising
models coupled together by products of the energy density
operators. This representation is convenient because the original
fermionic bilinears are local in terms of the Ising model order and
disorder parameter fields.  To clarify the symmetries and the
structure of the model, we can rewrite it as \bea
\label{effectiveH2}
H &=& H_{O(3)}[\chi;g_{\s,+}] +H_{O(2)} [\Theta_c^{(-)};g_{\rho,-}] +
H_{Ising}[\xi^3] \nonumber \\ &&+ 2\ri (\chi^a_R\chi^a_L)\{\ri
g_{\s -}(\xi^3_R\xi^3_L) + \bar g_{c,st}\cos[\beta\Theta_c^{(-)}]\}
\nonumber \\ && + \ri 2g_{c,ss}(\xi^3_R\xi^3_L)\cos[\beta\Theta_c^{(-)}].
%\label{int}.  
\eea 
Here $H_{Ising}$ is the Hamiltonian of the critical
Ising (CI) model and $H_{O(N)}$ represents the $O(N)$ Gross-Neveu (GN)
model \cite{GN} that can be described in terms of $N$ Majorana
fermions $\psi^a_{R,L}$ ($a=1,\ldots,N$) as \be
\label{GN}
H_{O(N)}[\psi;g_N]=- \frac{\ri }{2} v_N (\psi^a_R \p_x\psi^a_R -
\psi ^a_L \p_x\psi^a_L) - g_N(\psi_R^a\psi^a_L)^2 .  \ee The spectrum
of $H_{O(N)}$ is massive for $g_N < 0$, provided $N > 2$, while
$H_{O(2)}$ is always massless and is equivalent to the Gaussian model.

Though the form of the 
Hamiltonians (\ref{effectiveH}, \ref{effectiveH2}) is fixed by
the symmetry,  estimates of the coupling constants are available only
for weak interactions. The backscattering interactions are weak  in
such systems as carbon nanotubes which have the same symmetry as
two-leg ladders, but not in two-leg ladders themselves. Therefore  our
philosophy will be the same as in the particle physics: we will
express everything in terms of low energy parameters (mass gaps), and 
assume these gaps to be much smaller than the ultraviolet cut-off
(whatever this cut-off is).   As an additional
simplification we will ignore the difference between the velocities, 
setting $v_s = v_a = 1$.

\subsection{Preliminary RG analysis}

 The first step in our analysis is to establish the conditions under which the 
system scales to strong coupling. To do it we use the single loop 
RG equations as obtained 
in Ref.~\onlinecite{krice,Lin,Azaria,BF96}: 
\begin{subequations}
\label{RG}
\bea
&& \dot g_{\rho-} = -3 g^2_{c,st} - g^2_{c,ss}\\
&& \dot g_{c,ss} = - g_{\rho-}g_{c,ss} - 3g_{\s -}g_{c,st}\\ 
&& \dot g_{\s -} = - 2g_{\s +}g_{\s -} -  2g_{c,st}g_{c,ss}\\
&& \dot g_{c,st} = - (g_{\rho-} + 2g_{\s +})g_{c,st} - g_{\s -}g_{c,ss}\\
&& \dot g_{\s +} = - g_{\s +}^2 - g_{\s -}^2 - 2g^2_{c,st} \; .
\eea
\end{subequations}
Here the dot corresponds to a derivative in RG time $t = (4\pi
v)^{-1}\ln(\Lambda/\epsilon)$. Thus small energies in our notations
correspond to large $t$.  

Further in the text we will study the RG equations close to the
O(6)-symmetric point and, in more detail,  for  a simplified 
O(3)$\times$O(3)-symmetric model (cf. Sec.~\ref{sec:o3timeso3}). In
the latter case Eqs.(\ref{RG}) can be solved analytically.    

It is possible to show that  there are two areas of stability. 
\begin{itemize}
\item The weak coupling area,  where the repulsion dominates. All
  couplings scale to zero. The resulting phase is the O(6)$\times$U(1) TL
liquid perturbed by marginally irrelevant perturbations. We found the
exact boundaries of this area for the O(3)$\times$O(3) model, but
qualitatively one can say that the  TL liquid appears as a fixed
point  when the diagonal interactions are repulsive and exceed the
off-diagonal terms: $g_{\s+} \sim  g_{\rho,-} > |g_{\s-}|,
|g_{c,ss}|,|g_{c,st}|$.   
\item
The strong coupling area. In the RG sense this area is the basin of
attraction of the O(6)-symmetric point $- g_{\s+} = - g_{\rho,-} =
|g_{\s-}| = |g_{c,ss}| = |g_{c,st}| > 0$.   In order to get to strong
coupling it is enough to have just one attractive diagonal
interaction. Moreover and very intriguingly, one may even start from
all interactions being repulsive. The  system will scale to strong
coupling anyway provided the off-diagonal interactions are
sufficiently strong (this would correspond to emergent attraction
described in Introduction). This follows from the stability analysis
of the unstable symmetric line $g_{\s+} = g_{\rho,-} = |g_{\s-}| =
|g_{c,ss}| = |g_{c,st}| > 0$. 
\end{itemize}   

The RG equations determine  an overall scale at which the strong coupling is 
achieved:
\bea
M = \Lambda f[g_a(0)] = \Lambda \tilde f(g_1(0), C_1,... C_{Q-1}).
\eea
Each set of initial conditions sets the system on  a particular  RG
trajectory defined by its invariants $C_i (i = 1,...Q -1)$, where $Q$
is the number of the coupling constants. For a given set of $C_i$'s
the scaling limit exists if for an arbitrary large $\Lambda$ one can
choose a starting point on the trajectory such that $M$ remains
constant. It turns out that for trajectories with emergent attraction
this is not possible. Therefore  the space of $C_i$'s where the 
scaling limit is defined  is a subspace of $(Q-1)$-dimensional space 
(scaling subspace). Inside of this space the  masses may depend on 
$C_i$'s and this dependence is not determined perturbatively. In view
of this it is clear that the existence of  ``strong coupling point'' or
DSE requires independence (or at least weak dependence) of the mass
spectrum on RG invariants. Otherwise  strongly correlated phases are
not points, but have their own geography with  the excitation spectrum
changing  throughout the phase. We will continue this discussion in
Sections ~\ref{sec:1st} and \ref{sec:o3timeso3} where we consider mass
variations related to small deviations from the $O(6)$ and
$O(3)\times O(3)$ symmetries respectively.

\section{Strong Coupling Phases and Order parameters}
\label{sec:phases}

To explore the possible phases of the model  it is more convenient to
use the  mixed 
representation (\ref{effectiveH}). Here the Hamiltonian is expressed
in terms of bosonic 
fields $\Theta_{c}^{(\pm)}$ associated to the charge modes and Ising
fields $\sigma_a$ ($\mu_a$) with $a=0,\ldots,3$ associated to the spin
modes $\xi^3$ and $\chi^a$ respectively.
%one has to return to the bosonic formulation of model
%\ref{effectiveH}, 
%either in terms of the bosonic fields or in the Ising model form. 
The phases of the system are in one-to-one correspondence with 
 the fields vacua. These vacua are determined by the signs of the 
renormalized couplings ($g_{\s-}, g_{c,st}, g_{c,ss}$) in the strong 
coupling regime (recall that these signs may have nothing to do with
 signs of the bare coupling constants). The analysis of the phase
diagram  has already been conducted and here we  repeat many  results
obtained in Ref.~\onlinecite{Fradkin03,marston,TS04,Azaria}. 
To keep contact with these
works we use a strong tunneling approach introducing bonding ($p=1$) and
anti-bonding ($p=-1$) operators
\be
C(n)_{p,\s}=\frac{c_{n,1,\s}+p~c_{n,2,\s} }{\sqrt2}
\ee
associated to transverse wave vectors $q=0,\pi$ respectively. 
The bosonization notations are given in Appendix A and the
relationship with the weak tunneling approach is discussed at the end of
the section. 

When the forward scattering in the symmetric charge channel is not strong, 
there are four possible phases: 
superconducting s- (SCs) and d-wave (SCd), 
charge density wave (CDW),  and, what is now frequently called, d-wave CDW
(CDWd). Its order parameter is of Orbital Antiferromagnet 
\cite{OAF}-Staggered Flux \cite{affleck} type. In terms of the lattice
operators the order parameters of the four phases  have the form
\begin{subequations}
\bea
\label{scs}
&&\Delta_{SCs}(n)=\sum_{p =\pm 1} C(n)_{p,\s}C(n)_{p,-\s}  \nonumber
\\
&&= \sum_{p=\pm}\; [R_{p,\uparrow}L_{p,\downarrow} +
  L_{p,\uparrow}R_{p,\downarrow}] \\
\label{scd}
&&\Delta_{SCd}(n)=\sum_{p =\pm 1}  \sin(\pi
p/2)C(n)_{p,\s}C(n)_{p,-\s} 
\nonumber\\
&& =  \sum_{p=\pm}\; p \;
[R_{p,\uparrow}L_{p,\downarrow} + L_{p,\uparrow}R_{p,\downarrow}] 
\\
\label{cdw}
&&\Delta_{CDW}(n)=\sum_{p=\pm 1,\s} \;
C^\dagger(n)_{p,\s}C(n)_{-p,\s}\re^{ 2k_F \ri n a_0} \nonumber \\
&&= \sum_{p=\pm 1,\s}R^+_{p,\s}L_{-p,\s}\\
\label{cdwd}
&&\Delta_{CDWd}(n)=\sum_{p = \pm 1,\s} \; \sin(\pi p/2)
C^\dagger(n)_{p,\s}
C(n)_{-p,\s}\re^{2k_F\ri n a_0} \nonumber \\
&&= \sum_{p=\pm 1,\s}pR^+_{p,\s}L_{-p,\s}.
\eea
\end{subequations}
Here we  introduced the right and the left moving components of the
lattice fermion operators:
\bea
C_{p,\s}(n) = \re^{-\ri k_F^{(p)} x}R_{p,\s}(x) + 
\re^{\ri k_F^{(p)} x }L_{p,\s}(x), ~~ (x = na_0)
\eea
where $k_F^{(p)}$ are Fermi vectors of the different bands and 
$k_F=(k_F^{(+)}+k_F^{(-)})/2 = \pi(1 - \delta)/2a_0$,
with $\delta$ being the doping and $a_0$ is the lattice constant.
 
The SCd phase is found in the region $(+,+,+)$ and is characterized by 
\bea
\Theta_c^{(-)} =0, \la\s_a\ra \neq 0 (a=0,1,2,3) ~~ \mbox{or} ~~
\nonumber \\  \Theta_c^{(-)} =\sqrt\pi/2, \la\mu_a\ra \neq 0 (a=0,1,2,3)
\eea
where $\langle \ldots \rangle=\langle0|\ldots|0\rangle$ 
is the vacuum expectation value (VEV).
This phase has  power law correlations (quasi-long-range order) in 
the d-wave Cooper channel (\ref{scd}). The order parameter in the
continuum limit is 
\bea
\label{DSCd}
&&  \Delta_{SCd}    
\sim \re^{\ri\sqrt{\pi}\Theta_{c}^{(+)}} \\
&&\times \left\{\cos[\sqrt{\pi}
\Theta_{c}^{(-)}]\s_{1}\s_{2}\s_{3}\s_{0}  -
\ri\sin[\sqrt{\pi}\Theta_{c}^{(-)}]\mu_{1}\mu_{2}\mu_{3}\mu_{0}\right\}.
 \nonumber
\eea 

For  (--,+,--) one finds the CDWd phase, characterized by
\bea
&& \Theta_c^{(-)} =0, \la\s_a\ra \neq 0 ~~(a=1,2,3), \mu_0 \neq 0 ~~
\mbox{or} \nonumber\\
&&  \Theta_c^{(-)} =\sqrt\pi/2, \la\mu_a\ra \neq 0 ~~(a=1,2,3), \s_0 \neq 0.
\eea
One gets to this phase from the first one through a $\mathbb{Z}_2$
QCP.  In the continuum (\ref{cdwd}) becomes
\bea
\label{DCDWd}
&&  \Delta_{CDWd}  
\sim
\re^{-\ri\sqrt{\pi}\Phi_{c}^{(+)}} \\
&& \times \left\{\ri\sin[\sqrt{\pi}\Theta_{c}^{(-)}]
\mu_{1}\mu_{2}\mu_{3}\s_{0} -
\cos[\sqrt{\pi}\Theta_{c}^{(-)}]\s_{1}\s_{2}\s_{3}\mu_{0}\right\}. 
\nonumber 
\eea

From the CDWd going through a $U(1)$ QCP one reaches the 
CDW phase characterized by (--,--,+) and   
\bea
&& \Theta_c^{(-)} =\sqrt\pi/2, \la\s_a\ra \neq 0 (a=1,2,3), \mu_0 
\neq 0 ~~ \mbox{or} \nonumber\\
&& \Theta_c^{(-)} = 0, \la\mu_a\ra \neq 0 (a=1,2,3), \s_0 \neq 0.
\eea
The order parameter in the continuum limit takes the form
\bea
\label{DCDW}
&&  \Delta_{CDW}   
\sim \re^{\ri\sqrt{\pi}\Phi_{c}^{(+)}} \\
&& \times \left\{\cos[\sqrt{\pi}\Theta_{c}^{(-)}]
\mu_{1}\mu_{2}\mu_{3}\s_{0}- \sin[\sqrt{\pi}\Theta_{c}^{(-)}]
\s_{1}\s_{2}\s_{3}\mu_{0}\right\}. \nonumber 
\eea

\begin{figure}
%[ht]
\begin{center}
\epsfxsize=0.25\textwidth
\epsfbox{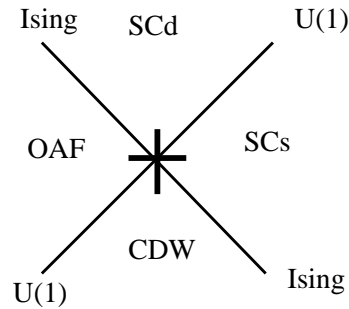}
\end{center}
\caption{The two-dimensional projection of the phase diagram for $K_c
> 1/2$. The thick lines represent first order phase transitions. At
$K_c < 1/2$ the SC phases are replaced by the Wigner crystal. } 
\label{fig:phases}
\end{figure}

Finally, the SCs phase is dominant in the region (+,--,--) and in
characterized by
\bea
&&\Theta_c^{(-)} =\sqrt\pi/2, \la\s_a\ra \neq 0 (a=0,1,2,3) ~~ \mbox{or}
\nonumber \\
&&  \Theta_c^{(-)} = 0, \la\mu_a\ra \neq 0 ~~ (a=0,1,2,3).
\eea
Its order parameter in the continuum limit is
\bea
\label{DSCs}
&&  \Delta_{SCs} 
 \sim \re^{\ri\sqrt{\pi}\Theta_{c}^{(+)}} \\
&& \times \left\{ -\ri\sin[\sqrt{\pi}
\Theta_{c}^{(-)}]\s_{1}\s_{2}\s_{3}\s_{0} +
\cos[\sqrt{\pi}\Theta_{c}^{(-)}] \mu_{1}\mu_{2}\mu_{3}\mu_{0}\right\}.
\nonumber 
\eea
This last phase can be reached both from the SCd phase through a $U(1)$ QCP
or from the CDW phase through a $\mathbb{Z}_2$ critical point.
All possible phase transitions are schematically shown in
Fig.~\ref{fig:phases} and can be summarized as follows: 
$\mathbb{Z}_2$ QCP between SCd and CDWd, SCs and CDW; U(1) QCP between
SCd and SCs, CDW and CDWd and  1st order transition between SCs and CDWd, and 
between SCd and CDW. More details about the phase boundaries are
provided in the next Section.

The scaling dimension of both CDWs phases is $d_{CDW} = K_c/4$, while
the one of SC phases is $d_{SC} = 1/4K_c$. The obtained bosonized
expressions for the CDW order parameters differ from the ones obtained
in Ref.s~\onlinecite{TS04,marston}. In our treatment of the order parameters we
used Ising model notations. This is done because Ising model order
and disorder parameters $\s$ and $\mu$ go naturally with Majorana fermions.

The density operator also contains a $4k_F = 2(k_{F}^{(+)} + k^{(-)}_{F})$ 
oscillatory piece
\bea
\rho(4k_F;q) =   \ri\exp[\ri\sqrt{4\pi}\Phi_c^{(+)}]
\left[A_q(\chi_R^a\chi_L^a) + B_q\xi_R^3\xi_L^3\right], 
\label{4kF}
\eea
where $q = 0,\pi$ is the transverse momentum. Notice that the
quasi-long- 
range order occurs only at the wave vector corresponding to the
overall particle density. Correlations with $4k^{(+)}_{F}, 4k^{(-)}_{F}$
turn out to be exponentially suppressed. 
The amplitudes $A_Q,B_Q$
vanish for the non-interacting system and  for weak interactions they are
of the  order of $U/\epsilon_F$. The importance of the $4k_F$
correlations was  pointed out in Ref.s~\onlinecite{giamarchi.book,marston}. 
Please notice that since the Majorana bilinears always have nonzero 
expectation values as soon as the spectrum is gapful,  the amplitude
of the $4k_F$ wave is finite except in the Tomonaga-Luttinger phase and
therefore critical 4$k_F$ density fluctuations are always present.
%(this fact was also discovered in \cite{marston}). 
As a consequence
the  $4k_F$ order parameter, having the  scaling dimension  $d_4 =
K_c$, competes with the SC ones. At $K_c < 1/2$ the scaling dimension
of (\ref{4kF}) becomes smaller than the scaling dimension of the SC
order parameters and SC phases on the phase diagram are  replaced by
the $4k_F$ CDW which we call Wigner crystal \cite{wigner}. It is possible that this mechanism is
responsible for charge ordering in the telephone number compound
observed in Ref.~\onlinecite{abba},\onlinecite{gorshunov},\onlinecite{abba2}.
As far as the CDW phases are concerned, the
$2k_F$ oscillations are always more relevant. However, an
establishment of a true long range 2$k_F$ CDW order in a
three-dimensional array of ladders will also lead to condensation of
the $4k_F$ oscillations. At special value of $4k_F = 2\pi$ (the 1/4
filled band) operator (\ref{4kF}) couples to the lattice. At this
point the charge field $\Phi_c^{(+)}$ couples to the other fields and
the corresponding sector acquires a gap. As follows from numerical
calculations done in Ref.~\onlinecite{WASc}, 
in two-leg ladders this may occur in a broad range of parameters.

 As far as  higher harmonics of the electron density  with the wave vectors $2nk_F$ are concerned, their scaling dimensions grow as $n^2K_c/4$ and they quickly become irrelevant. For this reason X-ray scattering from  such Wigner crystal shows sinusoidal oscillation of the electron density. This picture holds  when the dominant interactions are smaller than the one-dimensional bandwidth so that the continuous description can be used.    

The order parameters in the model with weak inter-chain tunneling 
(smaller than the gaps) can be obtained from the above order
parameters by the chiral particle-hole transformation:
\be
R_{l,\s} \rightarrow \s R^+_{l,-\s} \label{Dual}
\ee
Recall that in the transformed theory $l$ becomes a chain number and 
$\Phi$ and $\Theta$ fields in the charge sector interchange.

\section{Discrete symmetries and phase boundaries}
\label{sec:symmetries}

Apart from the continuous symmetries 
mentioned above, the Hamiltonian (\ref{effectiveH}) possesses 
a set of discrete symmetries. These symmetries establish a one-to-one
correspondence between excitation spectra of the different phases and
represent authomorphisms of the O(6) group \cite{Azaria}. To see this we
rewrite   the interaction (\ref{int}) in terms of 
SO$_1$(6) Kac-Moody currents $J_a = \psi_R\tau^a\psi_R$ and $\bar J_a = 
\psi_L\tau^a\psi_L$, where $\psi_{R,L}$ represent all six Majorana fermions
and  $\tau^a$ are generators of the O(6) symmetry group
\bea
V = G_{ab}J_a\bar J_b.
\eea
The discrete transformations which leave the Hamiltonian invariant
correspond to a change of sign of some chiral  
currents and the corresponding coupling constants, leaving the currents of
opposite chirality unchanged
\bea
J_a \rightarrow - J_a, ~~ \bar J_a \rightarrow  \bar J_a\;. \label{auto}
\eea
 Since the transformations (\ref{auto}) 
must preserve the commutation relations of the currents,  they correspond to  
authomorphisms of the corresponding group (the O(6) one in the given case). 
The above authomorphisms do not affect the spectrum, but change the order 
parameters (recall that the latter ones  are nonlocal in the
Majoranas) and  therefore 
establish a duality between different phases. 
%This fact was discovered by 
%Azaria {\it et al.} \cite{AZA}. 
In terms of Majorana fermions the  
authomorphisms corresponds to  a sign change of some  
Majorana fermions in one chiral sector. Therefore as far as the
spectrum is concerned, one can get a complete picture by studying it
in a given phase, for instance, in SCd. All other phases can be
obtained using the duality transformations. In the given 
case we have three dualities
\begin{itemize}
\item
$
\xi_L^3 \rightarrow - \xi_L^3, ~~ g_{\s,-}, g_{c,ss} 
\rightarrow - g_{\s,-}, - g_{c,ss}$
\item
$\chi_L^a \rightarrow - \chi_L^a, ~~ \xi_L^3 \rightarrow - 
\xi_L^3, \\ ~~~~~ g_{c,st}, g_{c,ss} \rightarrow - g_{c,st}, - g_{c,ss}$
\item
$\chi_L^a \rightarrow - \chi_L^a, ~~ g_{\s,-}, g_{c,st} \rightarrow - 
g_{\s,-}, - g_{c,st}$.
\end{itemize}

Let us now consider in more details the boundaries between the
 different phases.  
%From now on we concern ourselves with boundaries of strongly correlated 
% phases. 
Since the spectra of different phases are the same, we can take SCd
phase as  an example. 

The  boundary corresponding to a $\mathbb{Z}_2 $
QCP separating SCd phase from OAF is a
surface in  the coupling constant space on which the Majorana fermion
$\xi^3$  decouples from the rest and becomes 
massless. This surface is determined by the condition 
%If $\la(\chi_R^a\chi_L^a)\ra$ and
%$\la[(\xi^1_R\xi^1_L)+(\xi^2_R\xi^2_L)]\ra$ are non-vanishing and 
\bea
g_{\s,-}\la(\chi_R^a\chi_L^a)\ra +  g_{c,ss}\la[(\xi^1_R\xi^1_L) + 
(\xi^2_R\xi^2_L)]\ra =0 \label{cond1}
\eea
The criticality  is not 
violated by fluctuations around (\ref{cond1}) since they cannot generate any 
relevant operators. 

Analogously, if $(\chi_R^a\chi_L^a)$ and
$(\xi^3_R\xi^3_L)$ take expectation values and 
\bea
g_{c,st}\la(\chi_R^a\chi_L^a)\ra +  
g_{c,ss}\la(\xi^3_R\xi^3_L)\ra =0 \label{cond2},
\eea
the Majoranas $\xi^1, \xi^2$ decouple.  This is the U(1)
QCP between SCd and SCs. The fluctuations generate a perturbation 
$\cos[\sqrt{16\pi}\beta\Theta_c^{(-)}]$ which may become relevant if 
$\beta^2 < 1/4$. This, however,  corresponds  to a very strong $g_{\rho,-}$. Then this critical 
line may become a first order transition. 

Finally, if 
\bea
 g_{\s,-}\la(\xi_R^3\xi_L^3)\ra +  g_{c,st}\la[(\xi^1_R\xi^1_L) + 
(\xi^2_R\xi^2_L)]\ra =0 \label{cond3},
\eea
with $\la\xi^a_R \xi_L^a\ra$ different from zero,
the Majorana triplet $\chi^a$ is decoupled from $\xi$ fermions. At 
$g_{\s,+} <  0$ both models are massive,
 which means that this is a first order 
transition. The fluctuations shift $g_{\s,+}$ further to the negative side. If 
the effective coupling is positive, however, this becomes a $SU(2)_2$ quantum 
critical point. The first order line separates  SC and CDW phases with 
different point group symmetries (for instance, SCd and CDW).

For the analysis of the spectrum it is convenient to generalize the
model (\ref{effectiveH}) to $N$ species of fermions,
and rewrite the
interaction (\ref{int}) as 
\be
\label{form}
V = - \frac{1}{2}X^a \gamma_{ab} X^b \,;\,\,\, X^a = \ri\psi^a_R\psi^a_L .
\ee
The vicinity of the symmetric line of the phase diagram is described
by a weakly anisotropic $O(N)$ GN model
\be
\label{pert1}
\gamma_{ab}= \frac{1}{N}(g_0+\delta g_{ab})\,,\,\,\,\delta g_{a,b}/g_0 \ll 1.
\ee 
In an analogous way, at the phase boundaries the $N\times N$ matrix
$\gamma$
is split into two $N_1 \times N_1$ and $N_2 \times N_2$ blocks
$\gamma^1$ and $\gamma^2$ respectively
\be
\gamma^0 =
\begin{pmatrix}
\gamma^1& 0 \\
0 &\gamma^2
\end{pmatrix}
\,;\,\,\, \gamma^1_{a,b}=g_1\,,\, \gamma^2_{a,b}=g_2,
\ee
and the vicinity of the phase boundaries is characterized by
\be
\label{pert2}
\gamma_{a,b}=\gamma^0_{a,b} +\delta \tilde g_{a,b}
\,,\,\,\,\delta g_{a,b}/g_{1,2} \ll 1.
\ee
For the case at hand we have $N=6$, and 
\bea
&&N_1=5\, ,\,\,\,N_2=1, \,{\rm for} \,\,\,\mathbb{Z}_2 \\
&&N_1=4\, ,\,\,\,N_2=2, \,{\rm for} \,\,\,  U(1) \\
&&N_1=3\, ,\,\,\,N_2=3, \,{\rm for ~the ~first~ order}  
\eea
boundaries respectively.

%So, close to the unstable symmetric line 
%the the models take the form of an $O(N)$ GN model 
%perturbed by an anisotropy term 
%\be
%H_{eff}=H_{O(N)}[\psi;g]+\delta 
%g_{ab} \; \psi_R^a \psi_L^a \psi_R^b \psi_L^b
%\label{pert1}
%\ee
%with $\delta g_{a,b} \ll g_N$. Analogously, close to the phase
%boundaries it has either the form of two $O(N) \times O(M)$ GN models   
%with a weak coupling between the different groups 
%\be 
%\label{pert2}
%H_{eff}=H_{O(N)}[\chi;g]+H_{O(M)}[\xi;g']+
%\delta \tilde  g_{ab} \; \chi_R^a \chi_L^a \xi_R^b \xi_L^b.
%\ee
%or 
%\be
%H_{eff}=H_{O(N)}[\chi;g]+H_{Ising}[\xi]+
%\delta \tilde g \;\chi_R^a \chi_L^a \xi_R \xi_L.
%\label{pert3}
%\ee
%where the second GN model is substituted by a CI model.
%In what follows our main interest will be to study excitations of
%the above models in order to get insight into the spectrum of
%(\ref{effectiveH}) in the vicinity of the phase boundaries or highly
%symmetric lines.

\section{Semiclassical analysis}
\label{sec:semiclassical}
 In this Section we provide some general arguments concerning the spectra of 
generalized $O(N)$ GN models (\ref{form}-\ref{pert2}) based on the
semiclassical analysis valid in the limit of large
$N$ \cite{1/N}.  
A more intuitive analysis valid in the same approximation but  
for even $N,M$ is discussed in Appendix
\ref{app:semiclassical}. 

\subsection{Effect of small anisotropy on the vector particles}
\label{sec:sem1}

Let us consider first the  generalization (\ref{pert1}) of the GN model with a
total number of fermion species $N$ large. 
%in the limit when the anisotropy
%is small, and rewrite the interaction in (\ref{pert1}) as
%\be
%\label{form}
%V = - \frac{1}{2}X^a \gamma_{ab} X^b, 
%\ee
%with
%\be
%X^a = \ri\psi^a_R\psi^a_L,
%~~\gamma_{ab}= \frac{1}{N}(g_0+\delta g_{ab}),
%\ee
%where the coupling constants have been rescaled as standard procedure
%in $1/N$ analysis \cite{1/N} and $\delta g_{a,b}/g_0 \ll 1$.
If the quadratic form (\ref{form}) is negative defined, then it is possible to 
perform the 
Hubbard-Stratonovich transformation introducing an auxiliary field $\Delta^a$ 
\be
V=\frac{1}{2}\Delta ^a (\gamma^{-1})_{ab} \Delta^b -\Delta^a X^a,
\label{kink-fermion}
\ee
where $\Delta$ is related to $X$  by the equation of motion 
\be
\Delta^a=\gamma_{ab} X^b.
\label{deltax}
\ee
Integrating out the fermions one finds the effective potential  for $\Delta^a$
\be
V_{eff}[\Delta]=\frac{1}{2} \Delta ^a (\gamma^{-1})_{a,b} \Delta^b +
\frac{N}{8\pi} \Delta^2 \left ( \log \frac{\Delta^2}{\Lambda^2} -1 \right )
\label{veff},
\ee
that has minima at $\Delta^a=\pm m_a$ given by the saddle point equation
\bea
m_a = \frac{1}{2\pi}\sum_{b}\gamma_{ab} m_b\ln(\Lambda/|m_b|).
\label{fermionmasses}
\eea
From (\ref{deltax}) and (\ref{veff}) it follows that  
the fermion bilinears acquire vacuum  expectation values. 

Besides the trivial zero-energy solution, the classical static equation
of motion for the field $\Delta$ in the potential (\ref{veff}) also
has finite-energy, kink-like solutions, $\Delta_c(x)$, 
interpolating between different
minima (\ref{fermionmasses}). When quantizing the
theory \cite{GN,DHN} one finds that it
possesses fermionic excitations with masses $m_a$, 
associated to  the ordinary vacuum 
sector, and kink excitations, associated with  configurations 
$\Delta_c(x)$.
%fields interpolating between two different minima. 
In fact, as it was shown by Jakiw
and Rebbi \cite{jakiw.rebbi},  fermions interacting with topological
kinks like in (\ref{kink-fermion}) possess a single  normalizable zero
energy mode, $\psi_0$, in addition to the finite energy solutions
$\psi_n$. The crucial point is that, while finite energy solutions are
complex, the zero energy one is real and non-degenerate. The
semiclassical expansion is then given by \cite{jakiw.rebbi,witten}
\bea
\hat\psi^a(x,t) = \hat\gamma_a \psi_0^a+ 
\sum_n (\hat a_n \psi^a_n(x)\re^{- \ri t E_n} + h.c.), \label{jakiw}
\eea
where $\psi^0$ is the zero energy solution of the Dirac equation and,
assuming that $\Delta_c(+\infty) > 0$, it has the form 
\be
\psi_0^a=
\Omega^{-1/2} \left (
\begin{array}{c} 1\\0
\end{array}
\right)
\exp\left[ -\int_0^x \rd\xi \Delta_c^a(\xi)\right],\label{zeromode}
\ee
with $\Omega^{-1/2}$ being the normalization factor. 
Since $\psi_a$ are Majorana
fermions, the operators $\hat\gamma_a$ compose  a Clifford algebra
$\{\hat \gamma_a,\hat\gamma_b\} = \delta_{ab}$. Therefore these
matrices realize  spinor representations of the $O(N)$ group.
It is important to stress that the zero mode
solutions exist for any configuration $\Delta(x)$ which has
asymptotics of different sign (a kink).  
%These
%configurations have  nontrivial quantum numbers supplied by the zero
%modes of the Majorana  fermions bound to the kinks
%\cite{jakiw.rebbi,witten}
%\bea
%&& \hat\psi_a(x,t) = \hat\gamma_a\Omega^{-1/2}\left(
%\begin{array}{c} 1\\
%\ri
%\end{array}
%\right)
%\exp\left[ -e_a\int_0^x \rd\xi \Delta(\xi)\right] + \nonumber\\
%&& \sum_k \hat A_k \psi_k(x)\re^{- \ri t E_k}
%\eea
%It turns out that
%the fermionic excitations are kink bound states. 
The effect of small anisotropy on the kink masses is difficult to study in
the semiclassical limit; this problem  will be discussed in
Sec.~\ref{sec:1st} using other methods. We now consider in detail the
effect on fermions. 

The fermion masses 
are given by the saddle point equations
(\ref{fermionmasses}). 
The solutions of these equations 
depend on the bare couplings; for $\delta g_{ab}=0$
one recovers the usual form for the large $N$ limit GN mass, namely 
$m_a=M_0 = \Lambda\re^{- 2\pi /g_0}$. 
If $\delta g_{ab}$ is non vanishing but 
$|\delta g_{ab}|/g_0 << 1$,  the solution is 
\bea
\frac{\delta m_a}{M_0} =g_0^{-1}\left[\frac{1}{N}
\sum_{b \neq a}\delta g_{ab}\right] + g_0^{-2}\left[\frac{1}{N^2}
\sum_c\sum_{ b \neq c}\delta g_{cb}\right],  ~~ \label{masschange}
\eea
From this analysis it is clear that the vector multiplet is split, the  
splitting being 
proportional to the splitting of the bare coupling constants and 
survives  in the limit  $1/N \rightarrow 0$ provided $N^{-1}\sum
\delta  g_{ab} \neq 0$. 
 
Let us now imagine that some of the eigenvalues of $\gamma$ are
zero. The  limiting case is when the matrix $\gamma$ is proportional
to the projector:
\be 
\gamma_{ab} = \frac{1}{N}g e_a e_b, ~~ \sum_a e_a^2 =N \label{factor}
\ee
Then  Eq.(\ref{fermionmasses}) can be solved 
explicitly. The Hubbard-Stratonovich transformation yields  
\bea
{\cal L} = \frac{N}{2g} \Delta^2 + \left(\frac{\ri}{2}
\bar\psi^a\gamma_{\mu}\p_{\mu}\psi^a - \ri e_a\Delta\bar\psi^a\psi^a\right)
\eea
and  the  masses of the vector particles are given by 
\bea
m_a = e_a\la\Delta\ra
\eea
where 
\[
\la\Delta\ra = \Lambda \exp\left(- \frac{2\pi}{g} + A\right), 
~~ A = N^{-1}\sum_j e_j^2\ln (1/|e_j|)
\]
The $N \rightarrow \infty$ limit is defined as follows:
\bea
&&\sum_a = N\int \rd e \rho(e), ~~ \int \rd e e^2\rho(e) =1, \nonumber \\
&&A = \int \rd e \rho(e) e^2\ln(1/|e|) \nonumber 
\eea
So $N-1$ components of vector ${\bf e}$ are RG invariants.

\subsection{Kink confinement}
\label{sec:sem.confinement}

Let us now consider the opposite limit, when 
two $O(N_1)$ and $O(N_2)$ GN models ($N_{1,2}>2$) are
weakly coupled as in Eq.s~(\ref{form},\ref{pert2}). We discuss here only the
evolution of the kink excitations since the analysis for the fermions
is similar to the one carried out in the previous section.

In the absence of coupling, $\tilde g_{ab}=0$, 
each GN model has its own kinks, 
associated to two classical solutions, $\Delta_{1,2}(x)$, and  
interpolating between the minima $\pm m_{1,a}$ and $\pm m_{2,a}$  
respectively. 
The effect of the interaction is to lift the degeneracy
between the minima introducing a confining potential between the kinks. 
As a consequence the original kink solutions of the decoupled theories
become unstable and disappear from the spectrum. 

To see this let us consider,
for instance, the effect of the interaction
on a kink-antikink configuration, where both particles belong to the same 
GN theory, while the other theory is at the minimum, for example
$\Delta_2(x)=m_2$.  The kink-antikink configuration is such that 
$\Delta_1(x)$ takes value $-m_{1}$ from minus spatial infinity to a 
point 
$x_1(t)$ where it switches to $+ m_{1}$; it keeps this value until
$x_2(t)$ where it switches back to $-m_{1}$. In presence of the
perturbation, the kink-antikink state acquires an additional energy 
\be
U(x_1,x_2)=\delta\tilde g_{a,b} |x_1-x_2|m_1m_2. 
\label{univpot}
\ee
This  results in a confining potential for the
two $O(N_1)$ kinks (this universal form of the confining potential is
valid only when the confining radius is much greater than the size of
the lightest kink). For $N_1> 4$ this potential exists on top of the
attractive potential already present  in the $O(N_1)$ GN 
model. Therefore the  lowest bound states have the same symmetry as in
the $O(N_1)$ model, i.e. they transform according to the vector
representation of the group. Of course, if the confinement radius is much
greater than the inverse kink's mass, the confinement potential
contains many other bound states (we refer the reader to 
Sec.~\ref{sec:ffpt.confinement} for  a more detailed analysis). These
states do not exist in 
the $O(N_1 + N_2)$ GN model.
Their masses are grouped around the mass of the $O(N_1)$ vector
particle $m_{v1}$ or, if $N_1 \leq 4$, around $2M_1$, where $M_1$ is
the kink's mass in the $O(N_1)$ GN model. Repeating the same argument
for a kink-antikink configuration 
belonging 
to the $O(N_2)$ GM model one obtains the states grouped around the
mass of the $O(N_2)$ vector particle $m_{v2}$. 

The same confining phenomenon happens also between  two kinks that belong
to different GN models, in this case the lowest multiplet of the
confined states transforms according to the spinor representation of 
$O(N_1+N_2)$. This follows from the fact that such  states correspond
to a simultaneous change of sign of  $\Delta_1(x)$ and $\Delta_2(x)$,
therefore  they are kinks of the $O(N_1+N_2)$ GN model. 
Since the only interaction between kinks of the different models is
proportional to $\delta \tilde g _{a,b}$, the masses of these
particles group around the sum of the kinks masses $M_1 + M_2$.  The
details of the structure of the
excitations induced by the confining potential is discussed in
Sec.~\ref{sec:FFPT}.    
One should keep in mind that some multiple kink representations may survive 
(see Appendix~\ref{app:semiclassical}).

\subsection{Coupling massless and massive GN models}
\label{sec:massless}

The above analysis fails  if the matrix  $\gamma$  in Eq.~(\ref{form}) has
negative eigenvalues. If all  eigenvalues are negative, 
the analysis of the RG equations shows that all  interactions scale to zero
leaving  the theory massless.  The situation when the form has no
definite sign (at least, this is the criterion in the large $N$ limit) 
corresponds to the area  of emergent attraction. The large $N$
analysis is still applicable here, but in a modified form. Let us
consider the $O(N_1)\times O(N_2)$ model with $N_1/N_2$ finite in a
situation such that the interaction among the first $N_1$ particles is
attractive and the interaction among the other particles is
repulsive. We take a simple form of the interaction to keep the
analysis more clear. Then we can do the  Hubbard-Stratonovich transformation in
the attractive sector, so that after some algebraic manipulations we
obtain ($g_+,g_- > 0$): 
\bea
&-& \frac{g_+}{N_1}(\ri\bar\xi_a\xi_a)^2 +
2\frac{g_X}{N_1}(\ri\bar\xi_a\xi_a)(\ri\bar\chi_b\chi_b) +
\frac{g_-}{N_1}(\ri\bar\chi_b\chi_b)^2 \rightarrow \nonumber\\ 
&&  \frac{N_1}{2g_+}\Delta^2 + \ri\Delta \left (\bar\xi_a\xi_a -
\frac{g_X}{g_+}\bar\chi_b\chi_b \right ) + 
\nonumber\\ && 
N_1^{-1}\left (\frac{g_X^2}{g_+} + g_- \right
)(\ri\bar\chi_b\chi_b)^2.
\label{HS.massless}
\eea
Replacing $\Delta$ with a constant $M_0 \approx  m_{\xi}/g_+$ where
$m_{\xi}$ is a mass of $\xi$ fermions, we obtain the following
Lagrangian for $\chi$ fermions: 
\bea
{\cal L}  = \frac{\ri}{2}\bar\chi_a\gamma_{\mu}\p_{\mu}\chi_a + \ri
M_0\bar\chi_a\chi_a - g_{eff}(\bar\chi_a\chi_a)^2 \label{spin0} 
\eea
where $g_{eff} = N_1^{-1}(g_X^2/g_+ + g_-)$. Summing the diagrams
with a maximal number of loops, we obtain the following expression for
the mass ratio:  
\bea 
\frac{m_{\chi}}{m_{\xi}} = - \frac{g_X}{g_+}\left[1 +
  \frac{N_2}{N_1}(g_X^2/g^2_+ + g_-/g_+)\right]^{-1}. 
\eea
Thus also in this case the vector multiplet remains split. 

Let us now consider kinks. 
When  one of the GN models is massless the potential (\ref{univpot})
can no longer be used. Like above, we can do the Hubbard-Stratonovich
transformation in the massive sector (cf. Eq.~\ref{HS.massless}), and
study the problem of the two types of fermions in a background bosonic field.
For both fermionic modes the solution is again described by 
Eq.~(\ref{jakiw},\ref{zeromode}). As was discussed in Section
\ref{sec:sem1}, the entire   kink is a bound state of the scalar field 
$\Delta(x)$ and the fermions. The fermionic zero modes  now realize
spinor representations of the O$(N_1 + N_2)$ group and then in
principle can be seen as bound states of massive kinks of the $O(N_1)$
model and massless one of the $O(N_2)$.

The summary of these semiclassical arguments is the following. 
%illustrated by 
%Fig.~\ref{fig:spectrum1}. 
Weak  interaction between GN models generate (i) degenerate
kink multiplets transforming according to spinor representations of the 
$O(N_1 + N_2)$ groups and (ii) vector particles with different masses
transforming according to the vector representations of the $O(N_1)$
and $O(N_2)$ groups. On top of it there may be  many other multiplets
with masses laying between $(M_1 + M_2)$, $m_{v1}$, $m_{v2}$  
and $2(M_1 +M_2)$. 
These multiplets cross into the continuum and 
become progressively unstable once the
coupling between two models increases.

%where $M$ is the largest mass. This interaction leads to confinement of kinks and antikinks in both sectors. Let us restrict  ourselves to the  sector with two kinks (antikinks) and see what kind of bound states can emerge. Let us set $M_2 \geq M_1$. Then let both (anti)kinks belong to the same sector with a smaller mass. Then the interaction takes part in the background of static $\Delta_2$ and the effective action can be written as 
%\bea
%\ri m_0(\bar\chi^a\chi^a) + g_1(\bar\chi^a\chi^a)^2
%\eea
%where 
%\[
%m_0 \approx - \frac{\bar g}{g_1g_2 - \bar g^2}|_M\Delta_2
%\]
%This is a problem of vector particle formation in the O(N) sector. This problem has already been addressed in the previous subsection. The interaction with the other sector helps to form  vector particles. 

% The case not yet considered corresponds to the situation when  kinks belong to different sectors. Then  their bound states belong to spinor representation of the O(N +M) group.  Projecting energy (\ref{energy}) on the  two-kink state and restricting ourselves to small momenta, we arrive to the following Schroedinger equation in the center of mass frame:
%\bea
%\left[ - \frac{1}{2\mu}\frac{\rd^2}{\rd x^2} + \lambda |x|\right]\Psi(x) = (E - M_1 - M_2)\Psi(x)
%\eea
%where $x$ is the relative coordinate of the kinks, $\mu = M_1M_2/(M_1 + M_2)$ and $\lambda \approx 2g(M)\Delta_1\Delta_2$. This equation yields energy levels
%\bea
%E_n = (M_1 + M_2)\left[1 + \frac{1}{2M_1M_2}(\mu\lambda)^{2/3}\zeta(-n) + ...\right]
%\eea
%which gives approximately $g_0^2/\bar g$ stable energy levels.

\section{ Integrable points}
\label{sec:int.points}
At some specific points in the parameter space the Hamiltonian 
(\ref{effectiveH}) is
integrable. One can  identify the following integrable models, that 
describe either the most symmetric points or the phase boundaries.

{\em O(6) GN model}. As already noticed, if all coupling constants are equal
Eq.~(\ref{effectiveH}) has an extended $O(6)$ symmetry
and becomes the O(6)-symmetric GN model (\ref{GN}). The $O(N)$
GN model is integrable for any $N$ 
\cite{shankar-witten,Ann,OgReshW}.   
For even $N$ the scattering theory
is relatively simple \cite{Kar.Th}, while 
for $N$ odd there are significant complications (see
Ref.~\onlinecite{SalFend}). For $g<0$
and $N> 2$ the spectrum is massive and consists of kinks and
antikinks, which transform according to the irreducible spinor
representations  of the $O(N)$  group. For $N$ even kinks and
antikinks correspond to different representations, for $N$ odd there
is just one representation. 
The two-loop RG gives the following expression for the mass: 
\bea
M = \Lambda g^{1/(N -2)}\exp[- 2\pi/(N -2)g].
\eea
For $N> 4 $, there are  also 
fermion particles  and their bound states. These fermions  correspond
to the original Majorana fermions and transform according to the
vector representation
of the group. The masses of the fermions and their bound states  are given by
\be 
%\label{fermionmasses}
M_a=  2 M \sin \left ( \frac{\pi a}{N-2} \right ), 
~~a=1,\ldots,{\rm int}(\frac{N- 3}{2}).
\ee
For $N$ even, the total number of kinks is $2^{N/2}$ 
( reflecting the
fact that there are two $2^{N/2-1}$-dimensional  
irreducible spinor representations  of the algebra $SO(N)$
\cite{witten}). 
For $N$ odd
the total number of kinks is $2^{(N+1)/2}$ although there are major
subtleties associated to multi-particle states \cite{SalFend}. 
The number of fermions is always $N$.
In particular, for $N=6$ the spectrum consists of two spinor multiplets of mass
$M$ and one six-fold degenerate vector multiplet with mass $\sqrt{2} M$. No
fermion bound states are present. An intuitive picture of the two types of
excitations is provided in Appendix~\ref{app:semiclassical} for $N$
even. It is also good to remember that the spinor representations of
O(6) are isomorphic to fundamental representations of the SU(4)
group; their quantum numbers correspond to spin $\s =
\pm 1/2$ and transverse momentum $p = 0,\pi$ (or chain index for theories
with weak inter-chain tunneling).  

{\em $O(3)\times( U(1) \times \mathbb{Z}_2)$ model}. 
For $g_{\s-} = g_{c,st} =0$ two groups of Majoranas decouple from each
other; one is described 
by the O(3) GN model, the other one by the anisotropic O(3) GN 
(U(1)$\times$Z$_2$-symmetric). Both models are integrable (see
\onlinecite{nucTs,gold}), 
and the excitation spectrum contains only kinks.

{\em (3+1) model}. 
At the U(1) QCP, described earlier in Section III, two Majorana
fermions $\xi_{1,2}$ decouple and remain massless. The remaining
massive theory with the $O(3)\times \mathbb{Z}_2$ symmetry 
is related to an integrable model  solved by
Tsvelik \cite{Ts87} and Andrei, Jerez \cite{Andrei}. 
%\bea
%H_{TAJ}=\frac{2\pi v}{3} \left ( :J_1^2:+:J_2^2:+:\bar J_1^2:+:\bar J_2^2: 
%\right) \nonumber \\
%+\lambda (J_1 + J_2 )(\bar J_1 + \bar J_2) +\lambda '
%(J_1 - J_2 )(\bar J_1 - \bar J_2)
%\label{TAJ}
%\eea
(Some correlation functions for this model were calculated in 
Ref.~\onlinecite{ST03}.)
%The equivalence between  the two models follows from the following identities:
%\bea
%&& J_1=\frac{\ri}{2} \left [ \xi^3 \vec{\chi} + \frac{1}{2} 
%\left [\vec{\chi} \times  \vec{\chi} \right ] \right ] \\
%&& J_2=\frac{\ri}{2} \left [- 
%\xi^3 \vec{\chi} + \frac{1}{2} \left [\vec{\chi} \times
%  \vec{\chi} \right ] \right ], 
%\eea
%$\lambda=g_{\sigma,-}$ and $\lambda '=g_{\sigma,+}$.
The spectrum of this integrable model
contains an $SU(2)$ kink doublet with mass $M$ and a 
light Majorana fermion of mass $m_0\sim M(\lambda-\lambda')$.  
There are no stable vector particles except of this singlet fermion
who is, however, is {\it not} a bound state 
of kinks (see discussion at the end of the next paragraph).
The S-matrix can be found in App.~\ref{app:exactsol}.

{\em (5+1) model}. 
At the Ising QCP  one Majorana decouples. For $g_{\s,+}=g_{\rho,-}=g_{c,st}$
the massive model is  $O(5)$ symmetric and  integrable. 
%It turns out that it remains integrable 
Outside of QCP, provided the O(5) symmetry is maintained
($g_{c,ss}=g_{\s,-}$), it is possible to construct a factorized
scattering (integrable) theory with the same symmetry, $O(5)\times
\mathbb{Z}_2$, and UV central charge.  The exact solution is described in
Appendix~\ref{app:exactsol}; the spectrum consists of
two vector particles (quintet and singlet of Majorana fermions) with 
different masses $m_0$ and $\sqrt 3 M$ and a quartet  of kinks 
$\equiv$ antikinks with mass $M$ realizing the spinor representation 
of the O(5) group. This theory certainly deserves further analysis.
There are two serious  qualitative differences
between the scattering theories for the (5+1) model and  the O(6) GN
model (similar differences are found between the (3+1) model above and
the $O(4)$). 
First, as it follows from the group theory, in the former case
the kinks and antikinks are the same particles and in the latter case
they belong to  different representations. Second, in the (5+1)
theory the singlet fermion does not appear as a bound state of kinks
as all vector particles do in the O(6) GN model. As a consequence the
singlet  mass is not related to the kink's mass. One would imagine
however, that with increase of the coupling between the Majorana
singlet and the other ones the model approaches the O(6) GN model. At
present we do not have a complete picture of reconciliation of these
two models. The most likely solution to this puzzle is that the two
theories are equivalent only in the region where the singlet mass is small.

% How this happens is not clear at present. It is just possible, that
% the solutions
%ons of the (3+1) and (5+1) models derived in the
% assumption $m_0/M << 1$ \cite{%Ts87,Andrei} no longer describe the
% generalized GN model when $m_0/M \sim 1$. W%e admit that this problem
% requires a further study.  

\section{Perturbing  around integrable points}
\label{sec:FFPT}
 In this Section we discuss how to  calculate the spectrum using 
%other  methods than the $1/N$ expansion and quasi-classics. Namely, we shall 
%use 
perturbation theory around integrable points. Since integrable
points describe the phase boundaries and highly symmetric points, 
we will be able to obtain some additional 
 information about the excitations of the model in  these regions.

\subsection{Form-factor perturbation theory}
\label{sec:ffpt}

Let us first recall some essential features of integrable models that
we need in order to construct perturbation theory. 
Massive integrable models are characterized by a simplified on--shell dynamics 
which is encoded into a set of elastic and factorized scattering
amplitudes of their massive particles.  A convenient formalism for the
description of a dilute gas of particles with factorized scattering can be 
constructed in terms of the 
creation and annihilation operators, $A_a^\dagger(\theta)$,
$A_a(\theta)$, that satisfy  the Zamolodchikov-Faddeev (ZF)
algebra. (For an introduction on these concepts 
see for instance Ref.s~\onlinecite{Ann,book} or the introductory chapters of
Smirnov's book in  \onlinecite{smirnov}.) Here the
rapidity $\theta$ parameterizes the relativistic dispersion relation
\be
\label{e-p}
e_a(\theta)=M_a\cosh \theta, ~ p_a(\theta)=M_a\sinh \theta
\ee
and $a$ is an isotopic index. 
The ZF operators are the logical extension of the algebra of free fermions 
or bosons to the case of interacting particles with factorizable scattering, 
where the interaction is  completely characterized by the two-particle 
S-matrix, $S_{a,b}(\theta_{ij})$ \cite{Ann}. As usual, 
multi-particle states are  obtained acting with strings of creating
operators on the vacuum
\be
|\theta_1,\ldots,\theta_n\rangle_{a_1,\ldots,a_n} 
= A^\dagger_{a_1}(\theta_1) \ldots 
A^\dagger_{a_n}(\theta_n) |0\rangle.
\ee
The single particle states can be
thought as generated by an operator, 
$\varphi_a(x)$, such that $\langle 0 | \varphi_a |\theta \rangle_a \neq 0$. 
For the $O(N)$ GN model ZF operators include fermion,  
$A_{f_i}$ (and their bound states) and kink, $A_{k_i}$, operators.
Though for  the $O(N)$ GN model the S-matrix is not diagonal,
this will not be essential for what follows, so we prefer to describe the
methods for diagonal S-matrices since the notations are less cumbersome.

%In general, ZF operators are {\em non-local} 
%in terms of the original fermion  fields
%$c_{n,l,\s}$  and their local combinations. Therefore the physical  fields
%may create an arbitrary number of particles starting from some minimal number. 
 In what follows we will need a definition of non-locality. Recall  that two operators, ${\cal O}_1$ and ${\cal O}_2$,
are said to be mutually non-local if the Euclidean
correlator $\langle \ldots {\cal O}_1(x) {\cal O}_2 (0) \ldots
\rangle$ is not a single valued function of $x$. 
In particular, if we introduce 
complex variables $z=x^1+\ri ~x^2$ and $\bar z=x^1-\ri ~x^2$, and 
under analytic continuation $z\to z \re^{2\pi \ri}$,  
$\bar z\to \bar z \re^{-2\pi \ri}$, the correlator
acquires only a phase, $2\pi \gamma_{1,2}$, the
two operators are said to be {\em semi-local}. The non-local operators
that we will consider in the following are of this type.
%In what follows we restrict our consideration to {\em semi-local} operators. 
The great usefulness of this 
definition is that the index $\gamma_{1,2}$ can be calculated in the 
ultraviolet, where all correlation functions have a simple power law form.  
As we will see, the
effect of a perturbation on the spectrum of a theory will crucially
depend on the locality properties of the perturbing operator.

Let us imagine now that we  perturb an  integrable theory,
$H_{int}$, with a non-integrable 
perturbation 
\be
\label{Hperturbed}
H=H_{int} - g \int \rd x ~\Psi(x), 
\ee 
where $\Psi(x)$ is a scalar
field. % relevant (marginally relevant) operator with scaling
%dimension $\Delta_\Psi\leq 1$. 
The variation of the spectrum of the theory can be studied
perturbatively in $g$, in the same spirit as standard quantum
mechanics (QM) perturbation theory, taking advantage of the fact that
in integrable models matrix elements of perturbing operators 
\bea
&&F^\Psi_{b_1, \ldots, b_m; a_1,\ldots,a_n}(\theta_1', \ldots,
\theta_m'; \theta_1,\ldots,\theta_n) \nonumber \\ 
&&=~_{b_1, \ldots,
b_m} \langle \theta_1',\ldots,\theta_m' | \Psi(0) |
\theta_1,\ldots,\theta_n \rangle _{a_1,\ldots,a_n }, 
\eea 
or
form-factors (FFs), can be computed exactly \cite{smirnov}.  The
related perturbative approach is called Form-Factor perturbation
theory (FFPT) \cite{dms}.

According to Ref.~\onlinecite{dms}, 
the first order term in the expansion of the
mass variation of the particle $A_a$ is given by \be \delta m^2_a
\simeq 2 g~ _a\langle \theta | \Psi(0) |\theta \rangle_a= 2 g
\;C^{a,a'}\, F_{a' ,a}^\Psi(\ri\pi)\;,
\label{deltam}
\ee 
where the two-particle FF, $F^\Psi_{a_1,a_2}(\theta_1 -\theta_2)=
\langle 0 | \Psi(0) |\theta_1, \theta_2\rangle_{a_1,a_2}$ depends on
the difference of the rapidities only if $\Psi$ is a scalar
operator. In (\ref{deltam}) we used the crossing property valid for
Lorentz scalars 
%$_{a}\langle \theta_1 | \Psi(0) | \theta_2
%\rangle_a=\langle 0 | \Psi(0) | \theta_1+\ri\pi, \theta_2
%\rangle_{\bar a,a}$.  
%Please note that the kinks have non-trivial crossing and then this
%relationship has to be written in  a more general form
\bea
\label{gen.crossing}
&&F^\Psi_{b_1, \ldots, b_m; a_1,\ldots,a_n}(\theta_1', \ldots,
\theta_m'; \theta_1,\ldots,\theta_n) \\
&&=\prod_{j=1}^m C^{b_j,b_j'}
F^\Psi_{b_1', \ldots, b_m', a_1,\ldots,a_n}(\hat\theta_1', \ldots,
\hat \theta_m', \theta_1,\ldots,\theta_n) \nonumber 
\eea
where $\hat \theta=\theta+\ri \pi$ and 
$C$ is the charge conjugation matrix satisfying two
requirements: $C^t=C,\,C^2={\rm I}$. 
%Eq.~(\ref{deltam}) has to be
%modified accordingly. 
For Majorana fermions $C$ is trivial because they are neutral.

Since this is a strong coupling (IR) analysis,
if $g$ in (\ref{Hperturbed}) scales under RG, it has to be replaced in
(\ref{deltam}) by its renormalized value at energy of the order of the
largest mass in the theory 
\be
\label{geff}
g \to g^{eff} \simeq g(m).  
\ee 
The definition of the coupling constant at energy $m$ is 
somewhat ambiguous. The
problem is that  RG equations are universal only in the first loop and
beyond this they depend on the regularization scheme. By introducing
$g^{eff}$ we stretch these equations to their limit. Therefore we will
not be able to establish a rigorous relationship between the IR and UV
parameters.    

In analogy with QM degenerate
perturbation theory, the perturbed masses for degenerate multiplets
are obtained by diagonalizing the matrix $\{M_{m,n}
\}=\{F^\Psi_{n,m}(\ri\pi)\}$, where indices $n,m$ belong to the
degenerate multiplet. If the symmetry of the perturbing operator is
less than the symmetry of the multiplet, the perturbation will split
it.

  Assuming that the IR coupling constants in (\ref{deltam}) are smooth functions of the bare ones and  $F^\Psi(\ri\pi)_{\bar a, a}$
is finite, the spectrum evolves adiabatically.  However, there are
effects which do not appear in the first order.  The perturbation may
also generate an effective attraction between the particles, and lead
to an enlargement of the particle content of the theory through the
formation of bound states below the two-particle threshold. For small
$g^{eff}$ the appearance of bound states can be studied solving the
eigenvalue equation for the two-particle wave function. This will be
done in some detail later in the text (cf. (\ref{wf.mesons})).

A more dramatic change in the particle content of the theory happens
when $\Psi$ is non-local with respect to the operator, $\varphi_a$,
that generates the particle $a$ \cite{dms}. In this case, as shown
below, the two-particle FF, $F_{\bar a, a}^\Psi(\theta)$, has a pole
for $\theta=\ri \pi$ and then the mass variation (\ref{deltam})
diverges. This infinite mass variation implies that the original
particle $a$ (it will be called a ``quark'' in analogy with high
energy physics) is confined and disappears from the spectrum of the
perturbed theory. The resulting spectrum consists of quark bound
states - ``mesons''. Their masses may be higher than the two quark
threshold. The analysis of the meson spectrum will be done in
Sec.~\ref{sec:ffpt.confinement}.

The way the non-locality properties of an operator affect its FFs can
be seen as follows.  From the general theory of FFs it is known that
the $n$-particle FF has kinematical poles, associated to annihilation
processes, whenever the rapidities of a particle and anti-particle
differ by i$\pi$. The residue is given by \cite{smirnov} 
\bea
\label{poles}
&&-\ri {\rm Res}_{\theta'=\theta} F^\Psi _{\bar
  a,a,a_1,\ldots,a_n}(\theta'+\ri\pi,\theta,\theta_1,\ldots,\theta_n)=
  \\ && \left( 1-\re^{2\pi \ri \gamma_{a,\Psi}} \prod_{i=1}^n
  S_{a,a_i}(\theta-\theta_i) \right) F^\Psi _{
  a_1,\ldots,a_n}(\theta_1,\ldots,\theta_n), \nonumber 
\eea 
where $\gamma_{a,\Psi}$
is the semi-locality index between $\varphi_a$ and $\Psi$. It is
easy to see that for two particles ($n=0$) the right hand side
vanishes if $\gamma_{a,\Psi}$ is an integer. Then a two particle FF
has a pole at $\theta=\ri\pi$ only if the operator is non-local.  As
a consequence of (\ref{poles}) the mass variation (\ref{deltam})
induced by a non-local operator is infinite.
From (\ref{poles}) it follows that for $(\theta\sim \theta')$
\bea
&&F^\Psi_{\bar a,\bar
  b,a',b'}(\theta+\ri\pi,-\theta+\ri\pi,\theta',-\theta')
\simeq \nonumber \\&&-\alpha_{a,b} \langle \Psi\rangle
\left[ \frac{\delta_{a,a'} \delta_{b,b'}}{(\theta-\theta')^2} + 
\frac{\delta_{a,b'} \delta_{b,a'}}{(\theta+\theta')^2} S_{a,b}(2\theta')
\right ],  
\label{approx4FF}
\eea
with $\alpha_{a,b}=(1-\re^{2\ri\pi \gamma_{a,\Psi}})(1-\re^{2\ri\pi
\gamma_{b,\Psi}})$. This will be needed in the following.

The factorized scattering approach can also be constructed for
massless integrable models \cite{zam.massless}, despite the subtleties
in defining a scattering theory between massless particles in $(1+1)$
dimensions. In this case the excitations are right and left moving
particles, $A_{R}(\theta)$, $A_{L}(\theta)$, with dispersion relation
$e_R(\theta)=p_R(\theta)=\frac{M}{2}\re^{\theta}$ and
$e_L(\theta)=-p_L(\theta)=\frac{M}{2}\re^{-\theta}$, where $M$ here is
just a scale (for simplicity we
consider only particles with no internal indeces, this is the only
situation that we will encounter in the following). Also the FFs can
be defined and computed in analogy with the massive case
\cite{dms.massless}, and FFPT can be used to study the mass generation
that, to first order, will be given by \cite{cm} \be
\label{deltam.massless}
\delta m \simeq g~ F^\Psi_{R,L}(\ri\pi -\infty), \ee where
$F^\Psi_{R,L}(\theta)$ is the right-left FF \be
F^\Psi_{R,L}(\theta_{12})=\langle 0 |\Psi (0)| A_{R}^\dagger
(\theta_1) A_{L}^\dagger (\theta_2) \rangle .  \ee Also in this case
confinement is related to the non-locality properties of the
perturbing operator \cite{cm}.

\subsection{Non-locality properties of the perturbing operators}

In order to apply the above methods to deformations of GN models like
in (\ref{form}-\ref{pert2}), let us study the locality properties of
the perturbing operators with respect to the GN particles.

Consider first a massive $O(N)$ GN model ($N>2$).  Generally speaking
an operator of the form $\Psi_1=\psi^a_R \psi^a_L$ is local with
respect to the vector particles, $A_{f_i}$, but not with respect to
the kinks, $A_{k_i}$.  This can be easily seen even without explicitly
introducing kink creating operators.  In fact the kink is an
elementary excitation of the GN model that interpolates between
positive and negative minima of $\psi^a_R \psi^a_L$ and therefore this
operator changes sign across a kink configuration. This implies that
the non-locality index between the kink creating operator and $\Psi_1$
is $\re^{2 \ri \pi \gamma}=-1$, but, at the same time, an operator
that is a product of two fermionic bilinears belonging to the same GN
model, $\Psi_2=\psi^a_R \psi^a_L\psi^b_R \psi^b_L$, is local also with
respect to the kinks.  As a consequence, Eq.~(\ref{deltam}) can be
safely used to evaluate the first order effect of deforming the $O(N)$
GN according to (\ref{pert1}), since the perturbation is local with
respect to both kinks and fermions.  This implies that the spectrum
will evolve adiabatically, the only important effect is to split the
degeneracy between some states as discussed in the next section. 

On the other hand, if one takes two different massive GN models, say
$H_{O(N)}[\chi]$ and $H_{O(M)}[\xi]$ ($N,M>2$) and couples them like in
(\ref{pert2}), the perturbation, $\Psi_3=\chi^a_R \chi^a_L\xi^b_R
\xi^b_L$, is non-local with respect to the kinks of the two GN models,
thus leading to their confinement. This is in agreement with the
semiclassical analysis of Sec.~\ref{sec:semiclassical}.

\subsection{First order effects:
Perturbations around the $O(6)$-symmetric point}
\label{sec:1st}

We first apply FFPT to study the spectrum of vector particles around
the $O(6)$ symmetric point.  We consider small deviations of coupling
constants in (\ref{effectiveH}) from the point where all of them are 
equal $g_a = g + \delta g_a$
($a=[\rho,-],[c,ss],[\s,-],[c,st],[\s,+]$). 
The perturbation is local and then Eq.~(\ref{deltam}) can be
safely applied, with $g_a$ replaced by $g_a(m)$. It should be emphasized
that the coupling constants $g_a$ carry information about physical
quantities only when they are small. Beyond the first loop
approximation RG equations are not universal depending on the
regularization scheme. We assume that the interaction between physical
particles do not change substantially once $g_a \sim 1$ and therefore
stretching the RG equations to the limit of their  validity
$\ln(\epsilon/m) \sim 1$, we can estimate the anisotropy of the
coupling constants in the strong coupling regime. This assumption is
not justified but its consistency with know results will be checked.

For $\delta g _a/g \ll 1$ we can linearize the RG equations (\ref{RG}). 
The details of the calculations with all notations are given in 
Appendix \ref{appendix:lin}. We find
\bea
&&\delta g_a(\varepsilon) =
A_0\left[\log (\varepsilon/m) \right ]^{-2} +\label{var1}\\ 
&& \left ( T_{a,1} C_{-1/2,+}+T_{a,2} C_{-1/2,-} \right ) 
\left[\log (\varepsilon/m)\right] ^{1/2} \nonumber \\
&&+
\left ( T_{a,3}C_{1/2,+}+T_{a,4}C_{1/2,-} \right ) 
\left[\log (\varepsilon/m)\right ]^{-1/2} \nonumber
\eea
where $C_{q,\pm} \sim \delta_q g(0) g_0^{-q}$ 
($\delta_q$ stands for a particular linear combination of couplings)
are RG invariants, that can be obtained explicitly  using the results
of Appendix ~\ref{appendix:lin}, and
$A_0 \sim \sum \delta g_a(0) /g_0^2$. At scales of the order of the mass we
can set $\ln(\varepsilon/m) \approx 1$ in (\ref{var1}) and substitute the
result into  Eq.(\ref{deltam}). By dimensionality considerations the
matrix element of the current-current product is  const $m^2$. 
%where $C$ is
%a number. 
We remark that this normalization  yields the correct
estimate of the mass change at the uniform variation of the coupling
constants. Indeed, in  this case, according to (\ref{var1}), the
change of the effective coupling at energy $m$ is $\delta g_0/g_0^2$
and (\ref{deltam}) gives $\delta m \sim \delta g_0/g_0^2 m$. On the
other hand we obtain the same estimate  from the known dependence of
the mass on the bare coupling:    
\bea
\delta m^2 = \frac{\p m^2}{\p g_0}\delta g_0 = 
-\frac{2m^2\pi}{g_0^2}\delta g_0 \label{delmass}
\eea

 It is interesting to estimate the mass splittings for the case of
small $g_0$. Then the largest RG invariants that produce a
splitting of the multiplet are $C_{1/2}$ (the most relevant term,
proportional to $A_0$, is the same for any fermion). 
Taking only them into account 
we obtain  the following mass corrections:  
 \bea 
 && \frac{\delta m_c^{(-)}}{m} \sim
\left[\delta g_{\rho,-}(m)  + \delta g_{c,ss}(m) + 
 3\delta g_{c,st}(m) \right] \nonumber\\
&& \sim  C_{1/2,+} + C_{1/2,-} \nonumber \\
&&   \frac{\delta m_s^{(s)}}{m} \sim 
\left [3\delta g_{\s,-}(m) + 2\delta g_{c,ss}(m) \right ]  
\nonumber\\&& \sim C_{1/2,+} - 2C_{1/2,-}\nonumber \\ 
&& \frac{\delta m_s^{(tr)}}{m} \sim 
\left [ 2\delta g_{\s,+}(m) + \delta g_{\s,-}(m) + 
2\delta g_{c,st}(m) \right ] \nonumber\\
&&\sim - C_{1/2,+}\;.
\label{corrmass}
\eea
One can conclude from Eq.(\ref{corrmass}) that for any finite  RG
invariants the O(6) vector multiplet is split. Nevertheless the
splitting is very small in the regime of validity of this analysis.
%Obviously, in order to establish this splitting we From these
%equations it also follows that the splittings $m_s^{(tr)} -
%m_s^{(s)}$ and $m_c^{(-)} - m_s^{(s)}$ are independent of each other.  

Let us apply now Eq.~(\ref{deltam}) to study the evolution of the kink
masses. Since they are degenerate we need to diagonalize the matrix
with elements $_{k_i} \langle \theta_i \mid \psi^a_R \psi^a_L
\psi^b_R \psi^b_L \mid \theta_j \rangle_{k_j}$, where $k_{i,j}$
indicate kinks of the degenerate multiplet.  
The  needed FF has the form \cite{thanks}  
\bea
F_{k_i,k_j}^{a,b}(\theta_{ij}) &=& \langle 0 \mid \psi^a_R \psi^a_L
\psi^b_R \psi^b_L \mid \theta_i,\theta_j \rangle_{k_i,k_j}
\nonumber \\
&=&
(1-\delta_{a,b}) \, C_{k_i,k_j} f(\theta),
\label{kinkFF}
\eea
where $C$ is the charge conjugation matrix and the function 
$f(\theta)$ is finite at $\theta=\ri \pi$. 
From this, using (\ref{gen.crossing}), 
one gets that, for  $a\neq b$, 
\be
_{k_i} \langle \theta_i \mid \psi^a_R \psi^a_L \psi^b_R \psi^b_L 
\mid \theta_j \rangle_{k_j} = \, \delta_{k_i,k_j} f(\theta),
\ee
and then the matrix is diagonal. Using now 
Eq.~(\ref{deltam}) one finds that the mass variation is the same for
any kink and then the multiplet is not split.

\subsection{Beyond the first order}
\label{sec:ffpt.confinement}

\begin{figure}
%[ht]
\begin{center}
\epsfxsize=0.5\textwidth
\epsfbox{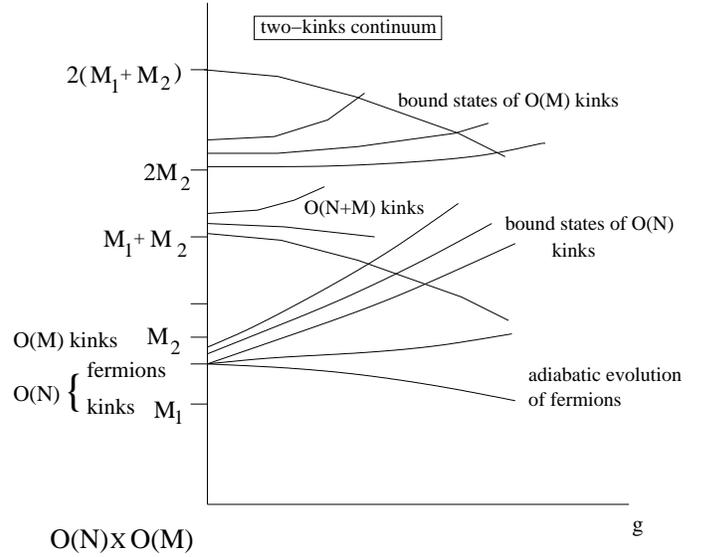}
\end{center}
\caption{A qualitative picture of the spectrum of the $O(N)\times O(M)$ 
GN model.}
\label{fig:spectrum1}
\end{figure}

Let us now turn to the problem of kink confinement induced by a
non-local operator. This analysis in important to study the
spectrum close to the first order transition line. 
In this case the FF in  (\ref{deltam}) is 
infinite and this formula cannot applied. Nevertheless,
for $g^{eff}$ sufficiently small the lower part of the meson spectrum 
can be studied within the two-quarks  approximation described in 
Ref.~\onlinecite{fonseca-zam} 
in the context of the Ising model. It is convenient to
parameterize the ZF operators directly in terms of momentum; the
relationship between the two parameterizations can be found in
(\ref{e-p}).
The two-particle wave
function can be written in the center of mass frame as
\be
\label{wf.mesons}
|\Psi_{a,b} \rangle =\int_{-\infty}^{+\infty} \rd
 \tilde\psi_{a,b}(p)~ A^\dagger _a(p) A^\dagger _b(-p)|0\rangle.
\ee
%with 
%\be
%\tilde\psi_{a,b}(\theta)= S_{a,b}(2\theta)\tilde\psi_{b,a}(-\theta) 
%\label{BC}.
%\ee
In general $a$ and $b$ can be different particles with masses $M_a$
and $M_b$, but both non-local with respect to $\Psi$. Please note that
the wave-function is not written in an explicit relativistic invariant form
although the states have a relativistic normalization.
The eigenvalue problem gives the following equation
\bea 
\left [ E- e_a(p)-e_b(p)\right ]
(\tilde\psi_{a,b}(p) \delta_{a,a'}
\delta_{b,b'}+\tilde\psi_{b,a}(p)
\delta_{a,b'} \delta_{b,a'} ) \nonumber \\
= g^{eff}
\int_{-\infty}^{\infty} \rd p' 
%F^\Psi_{\bar a,\bar
%  b,a',b'}(\theta+\ri\pi,-\theta+\ri\pi,\theta',-\theta')
%\tilde\psi_{a',b'}(\theta'),
~_{a,b}\langle p,-p \mid  \Psi(0) \mid p',-p' \rangle _{a',b'}~
\tilde\psi_{a',b'}(p'),
\label{sc1}
\eea 
where $g^{eff}$ is the renormalized coupling constant at energy 
$\sim (M_a + M_b)$. For energies close to the threshold (small
momentum) one can approximate 
$E- e_a(p)-e_b(p) \simeq \epsilon -p^2/2\mu$, with $\epsilon
=E-(M_a+M_b)$ and $\mu=M_aM_b/(M_a+M_b)$. If the 
operator is non-local,
it follows from (\ref{poles}) that, the right-hand side is dominated by the 
double pole in the four-particle FF.
It will be convenient for us to distinguish between two cases. One is
the  confinement of identical particles and the other is
when confining particles are not identical having, in general,
different masses. These two cases are qualitatively similar, but there
are differences in details. 

We consider first the confinement of two kinks that belong to
different GN models  induced by the operator $\Psi_3=\chi^a_R \chi^a_L\xi^b_R
\xi^b_L$. Let us call $a_i$ kinks that belong to 
$H_{O(N)}[\chi]$ and $b_i$ kinks that belong to
$H_{O(M)}[\xi]$. Given the structure
of the potential, the FF factorizes as
\bea
&&_{a_1,b_1} \langle p_1,p_2 | \Psi_3 | p_1', p_2'
\rangle_{a_2,b_2}  \\
&& = _{a_1}\langle p_1| (\chi_R^a \chi_L^a) | p_1'\rangle_{a_2} 
~_{b_1}\langle p_2| (\xi_R^b \xi_L^b) | p_2'\rangle_{b_2}. 
\nonumber 
%\\
%&&=F_{a_1,a_2}^{(\chi_R \chi_L)} (\theta_1-\theta_1'+\ri \pi)
%F_{b_1,b_2}^{(\xi_R \xi_L)} (\theta_2-\theta_2'+\ri \pi). \nonumber 
\eea
According to (\ref{poles}) and (\ref{approx4FF})
%, in the integral in (\ref{sc1}), 
for $p \sim p'$   we can approximate
\be
_{a_1}\langle p_1| (\chi_R \chi_L) | p_1'\rangle_{a_2} \simeq
%F_{a_1,a_2}^{(\chi_R \chi_L)} (\theta-\theta'+\ri \pi) \sim
\frac{2 \ri M_{a}}{(p_1-p_1')}  
\langle \chi_R \chi_L \rangle \delta_{a_1,a_2} \label{leading}
\ee
where we have used the fact that $\re^{ 2 \ri \pi
\gamma_{a,\Psi_3}}=-1$. Clearly the same approximation is valid for
%$ F_{b_1,b_2}^{(\xi_R \xi_L)} (\theta_2-\theta_2'+\ri \pi)$
$_{b_1}\langle p_2| (\xi_R \xi_L) | p_2'\rangle_{b_2}$.
From this we get  the eigenvalue equation  
\bea
(-\epsilon +p^2 /2\mu) \tilde\psi_{a,b}(p) = 
\lambda_{a,b}  ~ \mathcal{P} \int_{-\infty} ^\infty \frac{\rd p'}
{(p-p')^2}
\tilde\psi_{a,b}(p'),  
\label{sc-eq2}
\eea
were $\mathcal{P}$ indicates the principal part and 
\be
\lambda_{a,b} =4 g_{a,b}^{eff} \langle \chi_R \chi_L \rangle 
\langle \xi_R \xi_L
\rangle M_a M_b .\label{lambda}
\ee
After a Fourier transform to  configuration space  Eq.~(\ref{sc-eq2}) takes
the form
\be
(-\epsilon +\lambda_{a,b} 
|X|-\frac{1}{2 \mu}\frac{\rd^2}{\rd X^2} ) \tilde \psi_{a,b}(X)=0
\label{sc.realspace}
\ee
where
\be
\label{linear}
\tilde \psi_{a,b}(X)
=\int \frac{\rd p}{2\pi} \re^{\ri p X} \tilde \psi_{a,b}(p).
\ee
In general this equation must be supplemented by symmetry conditions.
%for the  wave function, which  is the zero rapidity limit of
%Eq.(\ref{BC}). However, if at $g^{eff} =0$ the theory decouples into
%two  independent models, like in this case,
%then kinks $a$ and $b$ are distinguishable
%particles with unity $S$-matrix. In that case  condition the (\ref{BC})
%becomes trivial.

The eigenvalue equation (\ref{linear}) describes a particle confined
in a  linear potential. The
solution is well known \cite{landau}. In the region $X>0$ one
can introduce the new variable  
$\zeta=(2\lambda_{a,b}\;\mu)^{1/3}(X-\epsilon/
\lambda_{a,b})$ so as to yield 
\be
(\frac{\rd^2}{\rd \zeta^2} -\zeta) \Psi(\zeta)=0
\ee
at $\zeta > -(2 \mu /\lambda_{a,b}^2)^{1/3} \epsilon$, 
from which it follows that the solution is the Airy function  
\be
\tilde\psi_{a,b}(\zeta)\propto {\rm Ai(\zeta)}.
\ee
 The eigenstates are determined  either by the zeros of the derivative
 of the  Airy function Ai$'(\zeta)$ at $X =0$ (symmetric wave
 functions), or   by the zeroes of the function itself (antisymmetric
 wave  functions). Calling them $\zeta_i$ we have 
\bea
&&E_i- (M_a+M_b) = ( \lambda_{a,b}^{2}/\mu)^{1/3}
\zeta_i~; \nonumber \\
&&\; ({\rm Ai}'(-\zeta_i)=0 ~~ {\rm or}~~ {\rm Ai}(-\zeta_i)=0).
\label{eigenvalues}
\eea
Clearly, if the particles $a,b$ are identical the above analysis
requires minor 
modifications as we will see below.

In agreement with the semiclassical arguments  of Section IV we see
that the kinks of the $O(N_1)$ and $O(N_2)$ models disappear from the
spectrum and are replaced by the generations of mesons which energies
belong to  the original two-particle 
continuum starting from $M_1+M_2$.  
Stable meson states must have spectral gaps smaller than 
$E_i<2E_1 \approx 2(M_1 + M_2)$. The above analysis is  quantitatively
valid if the number of meson generations is large. For the simplified
model $O(3)\times O(3)$ it is easy to  express this
number in terms of the RG invariants. 
Setting  $\Delta_{1,2} = $const$M_{1,2}$ we find from Eqs.~(\ref{C1C2})
below that at $g_+ \sim -1$ the effective coupling constant is   
\be
g^{eff} \sim [2|C_{1/2}| + C_{-1/2}]^{-2}
\ee
provided $|C_{1/2}| + C_{-1/2} >> 1$. 
Then  using  Eq.~(\ref{lambda}) we obtain
\bea
(g_{ab}^2/\mu)^{1/3} \sim (M_1 +
M_2)\left[g^{eff}\frac{\sqrt{M_1M_2}}{M_1 + M_2} \right]^{2/3} 
\eea
Since $\zeta_i \sim i^{2/3}$ at $i >> 1$, this sets the limit for the
number of meson generations as  
\be
i_{max} \sim g_{eff}^{-1}\left(\sqrt{M_1/M_2} + \sqrt{M_2/M_1}\right).
\ee

Thus it appears that the limit of weak confinement can be achieved by
either taking small $g_{eff}$ or considering quarks  with vastly
different masses. These two routes however lead to nonequivalent
limits. The limit $M_1/M_2 \rightarrow 0$ is different from the limit
$g_{eff} \rightarrow 0$ and cannot be studied using the above
formulae. The reason is that when the confinement radius
$(g_{eff}M_1M_2\mu)^{-1/3}$ becomes of order of the size of the
lightest kink $M_1^{-1}$, one can no longer use
Eq.(\ref{leading}). Thus the above analysis is valid only at  
\bea
M_1 >> g_{eff}M_2
\eea
At smaller mass ratios one can either use more accurate expressions
for the FFs or resort to the semiclassical analysis of Section
\ref{sec:sem.confinement}.  As we know from this  analysis,
confinement of a light particle on a heavy kink produces a zero energy
bound state and the scattering continuum separated from the bound
state by a gap of the order of $g_{eff}M_2$. This is of the order of
the energy of the lowest bound state $(g_{ab}^2/\mu)^{1/3}$ at $M_1
\sim g_{eff}M_2$.

Let us repeat the above analysis for confinement of identical
particles ($M_a=M_b=M$) induced, for instance, by an operator $\Psi$. As 
we discussed in Section IV, this leads to formation of vector particles.
Within the integral of (\ref{sc1}), we can
approximate
\bea
&&_{a,b}\langle p,-p \mid  \Psi(0) \mid p',-p' \rangle _{a',b'}
\simeq \nonumber \\&&- M^2 \langle \Psi\rangle
\left[ \frac{\delta_{a,a'} \delta_{b,b'}}{(p-p')^2} -
\frac{\delta_{a,b'} \delta_{b,a'}}{(p+p')^2} 
\right ]  .
\label{approx4ff}
\eea
With few manipulations Eq.~(\ref{sc1}) can be put in the form
(\ref{sc.realspace}) with $\lambda_{a,b}=4 g^{eff} M^2
\langle \Psi\rangle$. 
In this case the particles are indistinguishable so only antisymmetric
Airy functions are acceptable. The summary of the above results is
shown schematically in Fig.~\ref{fig:spectrum1}

It should be also noticed that if the unperturbed theory, like in the case
of the $O(N)$ GN model, possesses  bound
states of mass $M_{a,b}$ between particles $a$ and $b$ (associated with 
poles of the S-matrix on the physical strip), the eigenvalue equation
(\ref{sc1}) has to be modified taking into account this residual
attraction
\be
[ E- e_1(\theta)-e_2(\theta) ]  \to 
[ E- e_1(\theta)-e_2(\theta)  + \Delta E_{a,b}]
\ee
where $\Delta E_{a,b}= M_1+M_2- M_{a,b}$. As a consequence, 
Eq.~(\ref{sc.realspace}) is modified as 
\be
\left(-\epsilon -\frac{1}{2 \mu}\frac{\rd^2}{\rd X^2} + \lambda|X| 
- b_0\delta(X)\right)\psi_{\alpha,\beta}(X) =0 \;.
\ee  
The delta function is chosen in such a way that vector particles
appear also at $\lambda=0$. For the  $O(3)$ case, $b_0=0$.

The formation of bound-states induced by a local perturbation can be
discussed in similar terms. For small $g^{eff}$ the two-particle wave
function can be taken in the form (\ref{wf.mesons}) from which the
eigenvalue equation (\ref{sc1}) follows. The crucial difference with
respect to the case discussed above is that, since the operator is
local, the 4-particle FF in (\ref{sc1}) only have first order poles
giving rise to a non-singular potential.  Bound states form if the
perturbation is such that there is a solution of (\ref{sc1}) with
energies lower then $M_1+M_2$. This is what is found, for instance,
for the $O(4)$ GN perturbed by the $\mathbb{Z}_2$ anisotropy, as
discussed below.

\subsection{Coupling of massive and massless modes}
\label{sec:massless2}

The situation close to the $U(1)$ and $\mathbb{Z}_2$ critical points, 
described by Eq.~(\ref{pert2}) with $N_2=1,2$  is 
more tricky since some gapless modes are present.
This problem was already addressed in Sec.~\ref{sec:massless} and FFPT cannot
add much to that analysis. Nevertheless we find it instructive to
briefly discuss the problem also in this framework.
Let us look for simplicity 
at (\ref{pert2}) with $N_1=3,\, N_2=1$ and uniform deformation $\delta
g_{ab}=\delta g$.
For $\delta g =0$ the spectrum consists of a massless Majorana mode
and massive $O(3)$ kinks. 
According to (\ref{deltam}) and (\ref{deltam.massless})
the mass variation is given by
\bea
\label{deltam3}
&&\delta m_3 \simeq 2 \; \delta g \; \la \chi^a_R \chi^a_L\ra
 F^{(\xi^3_R \xi_L^3)}_{R ,L}(\ri\pi -\infty)\\
&&\delta m^2_\alpha \simeq 2 \; \delta g \;  \la \xi^3_R \xi^3_L\ra
\; _\alpha \langle \theta \mid \chi^a_R\chi^a_L \mid \theta \rangle
 _\alpha \; ,
\label{deltakink}
\eea
where the indices $3$ and $\alpha$ indicate  massless
Majorana fermions and massive $O(3)$ GN kinks respectively. The VEV 
$\la \chi^a_R \chi^a_L\ra$ is different from zero and
$F^{(\bar\xi^3 \xi^3)}_{R ,L}(\theta)$  
is constant for any value of $\theta$, then from (\ref{deltam3}) it 
follows that the perturbation induces a mass  
linear in $\delta g$, i.e. as soon as the coupling is turned on 
a massive singlet appears in the spectrum.

One needs to be careful in the analysis of
the second equation. In fact $\la\xi^3_R \xi^3_L\ra$ vanishes in the
unperturbed theory, implying that the mass variation of the kinks
vanishes in the  first order. Nevertheless $\la\xi^3_R \xi^3_L\ra\sim
\delta m_3$ and then (\ref{deltakink}) can still be applied, just
keeping in mind that the effect on the kinks mass is of higher order in 
$\delta g$. Since the operator $(\chi^a_R\chi^a_L)$ is non-local
with respect to the kinks,  the rhs of Eq.~(\ref{deltakink})
diverges and kinks confine.
One should notice that besides the one considered in (\ref{deltakink}), 
there are other
second order contributions to the mass variation of the kinks,
nevertheless they will not be able to compensate the divergence and
will not modify qualitatively the kink confinement.  It can be shown
that the confined states will always be unstable and then, in
agreement with the semiclassical analysis, the only stable particles
will be singlet and kinks with quantum numbers of the $O(N+1)$ model.  

\begin{figure}
%[ht]
\begin{center}
\epsfxsize=0.5\textwidth
\epsfbox{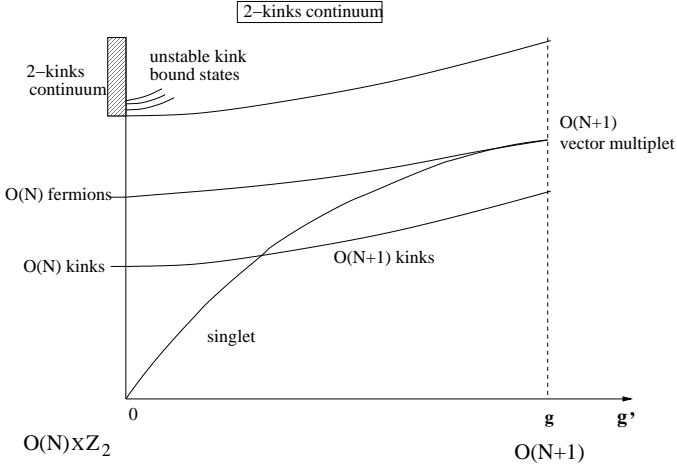}
\end{center}
\caption{A qualitative picture of the spectrum of the model
  interpolating between the $O(N)\times \mathbb{Z}_2$ and $O(N+1)$. 
GN model.}
\label{fig:spectrum2}
\end{figure}

These results can also be checked using the opposite limit $\delta g
\sim g_1=g_2$  and 
treating the model (\ref{pert2}) as a perturbation of the $O(4)$ GN
(remember that for $N=4$ the GN model has only kinks in the spectrum).
This perturbation is local with respect to the $O(4)$ 
kinks and then produces an adiabatic change of their masses. 
At the same time we know that the
$O(4)$ model also contain (unstable) vector particles above the two-kink
threshold. As described in the previous section  we can use FFPT to
check whether the perturbation induces further attraction that
stabilizes them.
Since the vector particles are bound states of kinks,
for small deviations from the $O(4)$ point we can write their wave
function like (\ref{wf.mesons}), with $a$ and $b$ being $O(4)$ kinks,
and study the stability of the bound state by
solving the eigenvalue equation (\ref{sc1}).
Even without entering the details of the computation one can easily
see that the form of the potential naturally split singlet and triplet
bound states and in particular the
interaction between kinks in the singlet state is attracting while the
one between kinks in the triplet state repulsive. Then it turns out that the
interaction stabilizes the singlet that will appear in the
spectrum.  This implies that the singlet state will be stable for any value of
$\delta g < g_{3}$ and will cross the threshold at the $O(4)$ point 
(see Fig.~\ref{fig:spectrum2}). 
We can repeat the same procedure for $N_1=5$, the
main difference is that the unperturbed theory has fermionic states in the
spectrum.

\section{The spectrum around phase boundaries}
\label{sec:spectrumatboundaries}
The analysis of the previous Sections allows us to understand
qualitatively, but rigorously, the
spectrum of the model (\ref{effectiveH})  close to the phase
boundaries, where it can be seen as a perturbation of some integrable models.

\subsection{From SCd to CDWd (SCs to CDW) phases through   Z$_2$ QCP}

At the $\mathbb{Z}_2$ QCP  the  Hamiltonian (\ref{effectiveH}) is
described by a deformed $O(5)$ GN model decoupled from the Critical Ising 
model. The effect of the anisotropy is to lift the degeneracy between
the vector particles. Away from the $\mathbb{Z}_2$ QCP the coupling is 
\bea
 - 2(\xi^3_R\xi^3_L)\left\{g_{\s-}(\chi^a_R\chi^a_L) + 
g_{c,ss}[(\xi^1_R\xi^1_L) + (\xi^2_R\xi^2_L)]\right\} .
\eea
It is convenient to study the spectrum for 
$g_{\s,+}=g_{\rho,-}=g_{c,st} = g; g_{\s-} = g_{c,ss} = g_X$ when the
$O(5)$ symmetry is exact. The RG equations 
\bea
\dot g = - 3g^2 - g_X^2, ~~ \dot g_X = - 4g_Xg
\eea
have the following RG invariant: $C = (g^2 - g_X^2)/g_X^{3/2}$. 
To understand the spectrum we can use results from the previous
sections as well as the exact solution of the $(5+1)$ model presented
in App.~\ref{app:exactsol}. 
The spectrum consisting of 5+1 vector particle multiplet and a
quartet of kinks.  The symmetry of the spectrum remains 
O(5)$\times$Z$_2$, which corresponds to the splitting of the vector
multiplet as 5 + 1. Following the arguments of Section \ref{sec:1st}
we find that at large $C >> 1$ the mass ratio of the different vector
particles is $\sim g_X(m) \sim C^{-2/3}$. At $C << 1$ the vector
multiplet approaches the O(6) symmetry with the splitting  $\sim C$.   

Then in general, the SCd phase has kink particles with the $O(6)$ quantum
numbers and vector multiplet containing a triplet,
a doublet and a singlet with different masses. All masses have the
same sign. By  going trough Z$_2$ QCP the singlet mass changes sign
and one finds the Orbital Antiferromagnet (CDWd) phase.

\subsection{From one SC (CDW) phase to another one  trough  U(1) QCP} 
 
At the $U(1)$ transition two Majorana fermions are massless and the rest
is a deformed $O(3)$ GN model equivalent, for small deformations, 
to the (3+1)-model discussed in
Sec.~\ref{sec:int.points}. Thus the spectrum consists of  the Gaussian
massless mode, the $SU(2)$ 
kink doublet  and a singlet particle of a lower mass. 
The  $\Theta_c^{(-)}$ field acquires a mass gap as soon as the coupling 
between the two sectors is on.  
The effective field theory describing the 
lowest excitations is the sine-Gordon (SG) one:
\be
S = \frac{1}{2}[\p_{\mu}\Theta_c^{(-)}]^2 + v\cos[\beta\Theta_c^{(-)}], 
\ee
with $v = \ri\la[g_{s,ct}(\chi^a_R\chi^a_L) + 
g_{c,ss}{\xi_R^3\xi_L^3}]\ra $ and $\beta^2 = 4\pi(1 - g_{\rho,-}/2\pi)^{-1}$.
The SG kinks with period $2\pi/\beta$  constitute a low-lying
doublet.  Thus the vector multiplet is reduced to the U(1) doublet
and the Majorana singlet. For $\beta^2 > 4\pi$ (repulsive
interactions) there are no sine-Gordon bound states, but if  
$\beta^2 < 4\pi$, are also additional bound states (breathers).  
These particles are absent in the
O(6) GN model.  As far as the kinks are concerned, they acquire both
spin and U(1) quantum numbers via creation of bound states between
half (anti)kinks of $\Theta_c^{(-)}$ with period $\pi/\beta$ and  the
SU(2) kinks.

\subsection{ Around the first order line}
At the first order transition the spectrum consists of kinks with different
masses, $M_1$ and $M_2$
belonging to the isotropic and anisotropic $(U(1)\times \mathbb{Z}_2$)
O(3) GN models respectively. 
As soon as the interaction is turned on all kinks confine and
one has fermions close to the two kink thresholds $2M_1$ and $2M_2$ and new
kinks close to the $M_1+M_2$ threshold. As we discussed earlier in 
Sec.~\ref{sec:ffpt}, the region of weak confinement where one may 
observe several generations of kinks, is likely to be narrow.

\section{Further from the boundaries:
O(3)$\times$O(3) model as a simplified case. }
\label{sec:o3timeso3}

\begin{figure}
%[ht]
\vspace{5mm}
\begin{center}
\epsfxsize=0.4\textwidth
\epsfbox{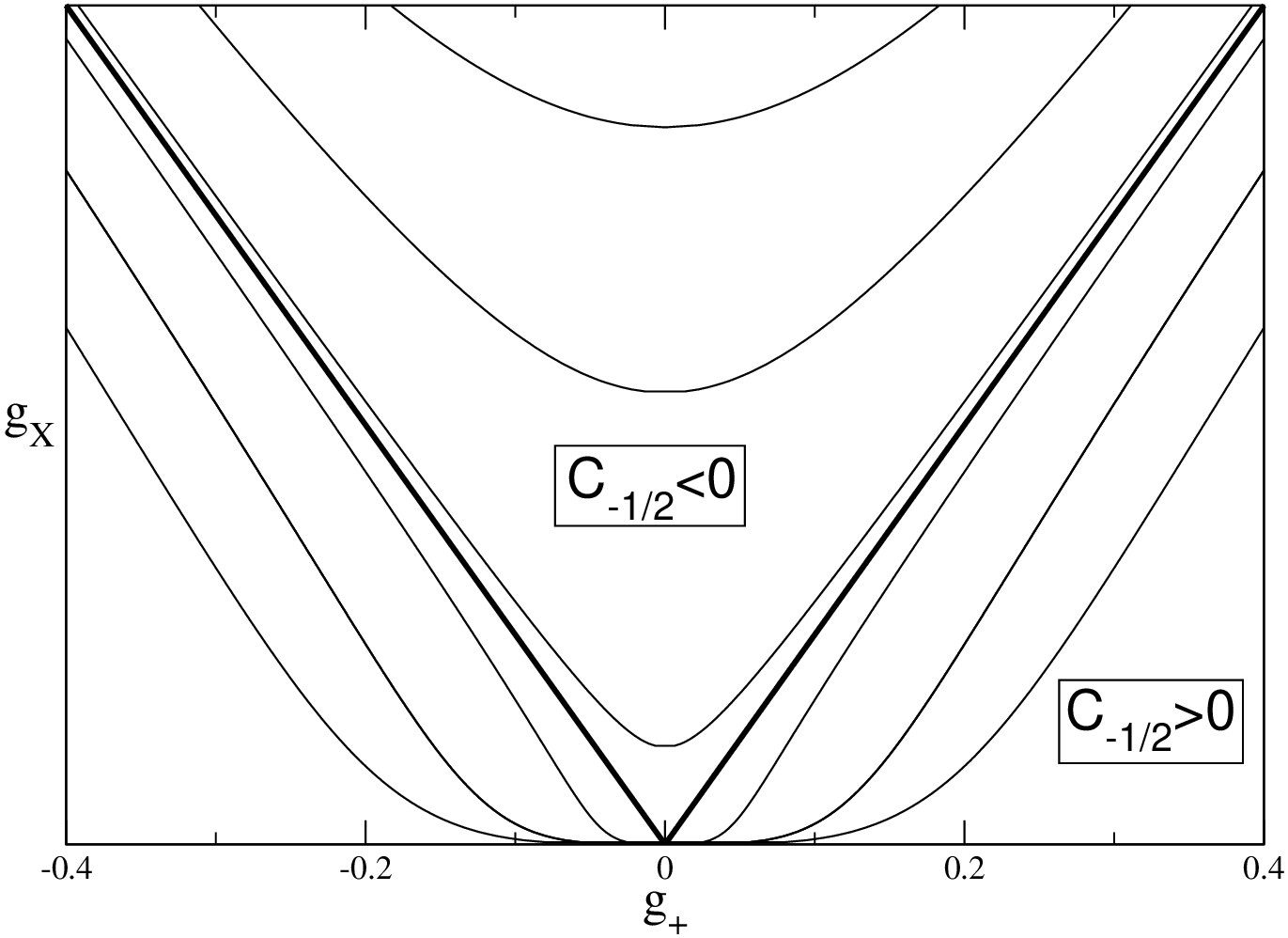}
\end{center}
\caption{RG flows (\ref{RG2}) for $C_{1/2} =0$. The thick line represents
  the flow with $C_{-1/2} =0$. } 
\label{fig:RG1}
\end{figure}

With the results of the previous sections at hand we will return to
the case  $N=6$ and consider the model (\ref{effectiveH}) with a more
symmetric form of the interaction 
\bea
&&  
V = - g_{+}(\chi^a_R\chi^a_L)^2 
\nonumber\\
&& 
- 2g_{X}(\chi^a_R\chi^a_L)(\xi^b_R\xi^b_L)  - 
 g_{-}(\xi^a_R\xi^a_L)^2.
\eea 
as it was done in ~\onlinecite{1D,Azaria.o3o3,IQA}. Though the  subsequent RG analysis  has a significant overlap with the one conducted in these papers, we include it here for completeness. The RG equations have the following form:
\bea
&& \dot g_+ = g_+^2 + 3g_X^2, ~~ \dot g_- = g_-^2 + 3g_X^2, \nonumber\\
&& \dot g_X = 2g_X(g_+ + g_-), \label{RG2}
\eea
In this case we have managed to find  the RG invariants explicitely
\bea
C_{-1/2} = \frac{g_+g_- - g_X^2}{\sqrt{|g_X|}}, ~~ C_{1/2} 
= \frac{g_+ - g_-}{2\sqrt{|g_X|}}. \label{C1C2}
\eea
For $g_a = g + \delta g_a$ these  invariants correspond to the ones
found in Section \ref{sec:1st}. If the O(3)$\times$O(3) model is
replaced by O(N)$\times$O(N) the power 1/2 in (\ref{C1C2}) is replaced
by $(N-2)/(N-1)$. This means   that  in the large $N$ limit the ratios  of 
coupling constants become RG invariants which coincides with
conclusions of Section \ref{sec:semiclassical}.  We also observe  that
a sign of the quadratic form $\gamma$ is RG invariant  which gives
support to the $1/N$ analysis of Section
\ref{sec:semiclassical}. Using these RG invariants one can integrate
Eqs.(\ref{RG2}). It should be  kept in mind however, that equations
(\ref{RG2}) together with the explicit form  (\ref{C1C2}) are valid
only at $|g_a| < 1$.  As soon as one of the couplings becomes $\sim 1$
the scaling of this coupling should be stopped and the problem
reconsidered.

\begin{figure}
%[ht]
\vspace{5mm}
\begin{center}
\epsfxsize=0.4\textwidth
\epsfbox{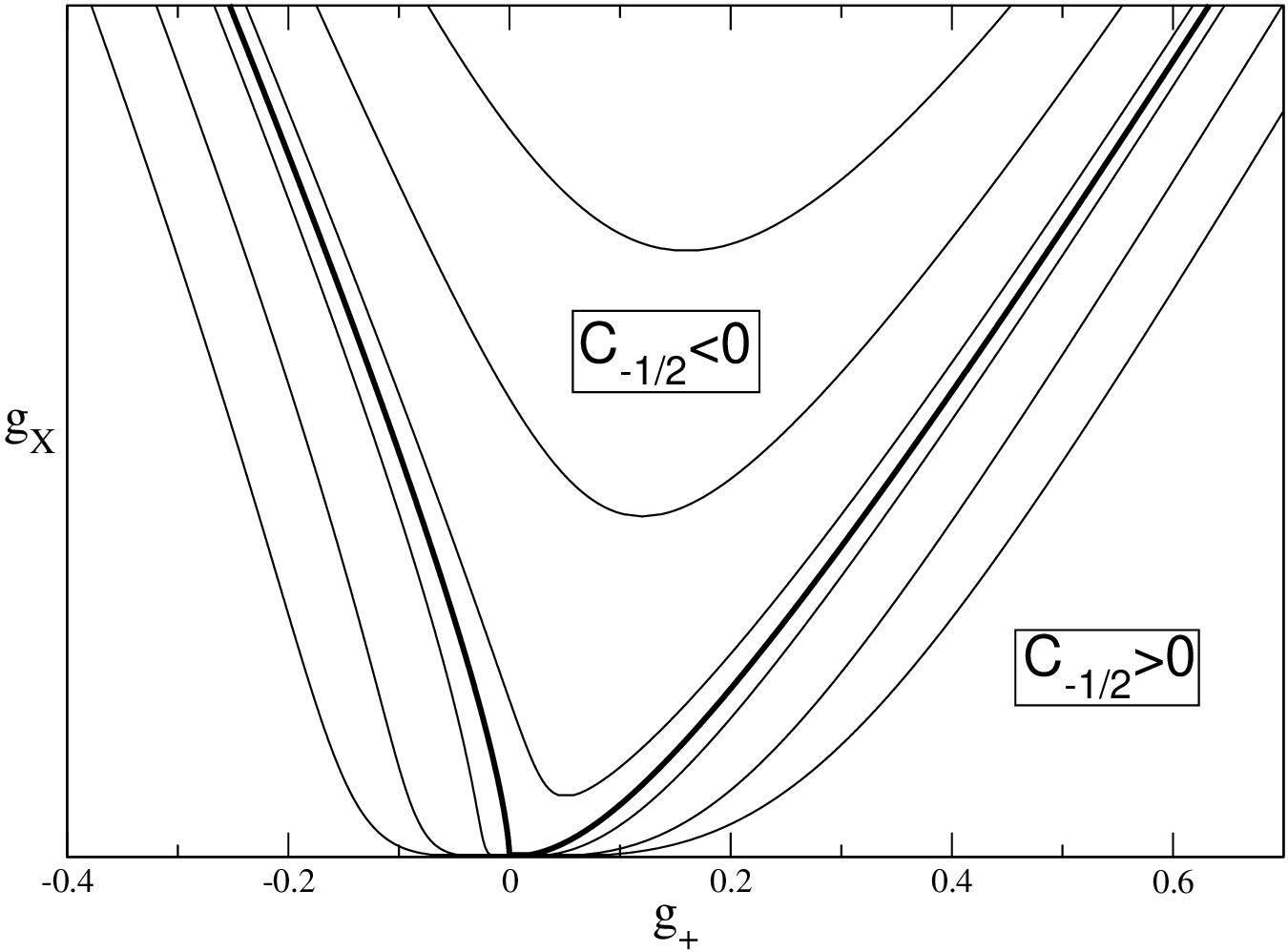}
\end{center}
\caption{RG flows (\ref{RG2}) for $C_{1/2} > 0$. The thick line
  represents  the flow with $C_{-1/2} =0$. } 
\label{fig:RG2}
\end{figure}

The solution of this equation is given by 
\bea
 4t = \pm \int_{g_X(0)}^{g_X(t)} \frac{\rd g}{g\sqrt{C_{-1/2}g^{1/2} +
     C_{1/2}^2g + g^2}}\label{solRG}, 
\eea
that can be expressed in terms of elliptic functions of complex
modulus and argument.

The phase diagram has three areas corresponding to different
properties of the interaction tensor. 
\begin{itemize}
\item
The signature of the interaction tensor is positive: $C_{1/2} > 0$ and
$C_{-1/2} > 0$. All  couplings scale to zero. 
\item
The signature is negative ($C_{1/2} < 0$) and  $C_{-1/2} >0$. The
system scales to strong coupling. The energy scale on which the
spectral gaps are formed is given by  
\bea
 M \sim \Lambda\exp\left[ -\pi\int_{|g_X(0)|}^{\infty}\frac{\rd
     g}{g\sqrt{C_{-1/2}g^{1/2} + C_{1/2}^2g + g^2}}\right]
 \label{Mass} 
%&& \Lambda\exp\left[ -\frac{\pi}{|g_X(0)|}\int_{1}^{\infty}\frac{\rd
%x}{x\sqrt{%[C_1/g_X(0)^{3/2}]x^{1/2} + [C_2^2/|g_X(0)|]x +
%x^2}}\right] \label{Mass} 
\eea
At $C_{-1/2} >0$ the integral in the exponent diverges at $g_X(0)
\rightarrow 0$.  Therefore the theory has a proper scaling limit $M =
$const $g_X(0) \rightarrow 0, \Lambda \rightarrow \infty$. This limit
is characterized by the RG invariants $C_{\pm 1/2}$. The case $C_{\pm
  1/2} = 0$  corresponds to the O(6) GN model. The opposite limit
max$(C_{\pm 1/2}) >> 1$ describes  the O(3)$\times$O(3) model in the
regime of weak confinement   
(see Sections \ref{sec:sem.confinement},\ref{sec:ffpt.confinement}).

\item
$C_{-1/2} <0$. The system scales to strong coupling for any sign of
  the bare diagonal couplings. In this case the scaling trajectory for
  $g_X(t)$ bounces off its minimal value $g*$ given by the root of the
  equation  
\be
[g^*]^{3/2} + C_{1/2}^2[g^*]^{1/2} - |C_{-1/2}| =0 \label{g*}
\ee
 If $g_X(0) > 0$ the weakest interaction is achieved at the energy scale 
\bea
E^* = \Lambda\exp\left[ -\pi\int_{g^*}^{g_X(0)}\frac{\rd
    g}{g\sqrt{C_{-1/2}g^{1/2} + C_{1/2}^2g + g^2}}\right] 
\eea
 and the strong coupling regime is reached at 
\be
M = E^*\exp\left[ -\pi\int_{g^*}^{\infty}\frac{\rd
    g}{g\sqrt{C_{-1/2}g^{1/2} + C_{1/2}^2g + g^2}}\right]. 
\ee
As we said in Introduction, the maximal spectral gap 
corresponding to  $g_X(0) \rightarrow +\infty$ is 
\bea
M \sim {E^*}^2/\Lambda
\eea
 Since integral (\ref{solRG}) never diverges at $C_{-1/2} < 0$, the RG
  time in which the strong coupling limit is reached always remains
  finite. Therefore at finite $C_{-1/2} <0$ the  scaling limit does
  not exist.   
\end{itemize}

If we relax conditions on the scaling limit we can venture into the area 
$C_{-1/2} < 0$, as far as the spectral gaps remain much smaller than
the bandwidth.  For this we need $g^*$ to be small which is achieved
by making $C_1$ small. Thus the effective low energy theory
describing the state with emergent attraction is  the  $C_{-1/2} =0$
field theory  with non-relativistic corrections in  $M/\Lambda$. The
condition $C_{-1/2} =0$ does not put any restrictions on $C_{1/2}$;
in the realistic two-leg  ladder, where bare lattice interactions are
not small, $C_{1/2}$ may have any value.  Eq.(\ref{C1C2}) pointing to
small values of $C_{\pm 1/2}$ at small bare couplings is not valid in
that limit.  

\begin{figure}
%[ht]
\begin{center}
\epsfxsize=0.45\textwidth
\epsfbox{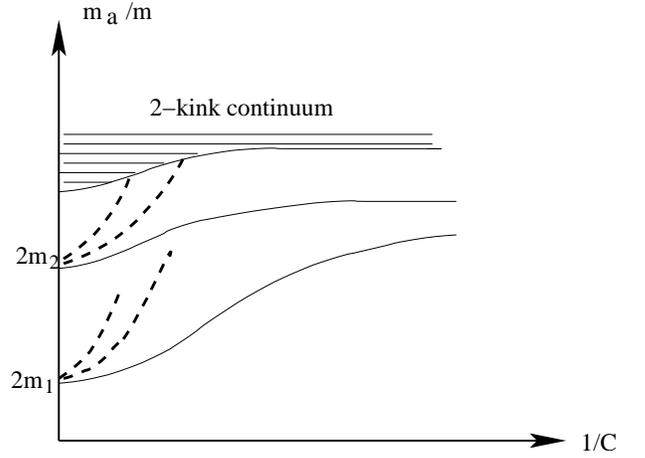}
\end{center}
\caption{The qualitative picture of the mass spectrum of the
  O(3)$\times$O(3) GN model as a function  $1/C_{1/2}$. The dashed
  lines show masses of mesons. }    
\label{fig:mesons}
\end{figure}

Taking the above into account we can qualitatively describe the
overall spectrum of the O(3)$\times$O(3) using results of Sections
\ref{sec:1st},\ref{sec:ffpt.confinement}. The results are summarized
on Fig. \ref{fig:mesons}.

\subsection{The mass is induced in the otherwise  massless sector}

 It remains to study the situation when some coupling constants
are much larger than the others such that they reach the strong
coupling limit first. This analysis is complementary to  the one of
Sec.s~\ref{sec:massless} and \ref{sec:massless2}.  
In this case one should study  the
spectrum using two-stage RG. There are two cases to consider: 
\begin{itemize}
\item
One of the sectors is massless at $g_X =0$. Let it be the $+$ sector:
$g_{-}(0) < 0, g_{+} > 0$, so that at $g_X =0$ the interaction in
$\chi_a$ sector is marginally irrelevant. The condition $C_{-1/2} < 0$
is fulfilled.  As we shall demonstrate,  the vector particles do
appear in this sector at $g_X \neq 0$, but the symmetry is not
restored. 
\item
Both sectors are massive at $g_X =0$. Let  $0< |g_X(0)| <<
g+(0)g_-(0)$ and $-g_{+}(0) \sim - g_{-}(0) > 0$. Then the coupling
$g_X(M)$ is still weak. Following the arguments of the previous sections,
we conclude that the interaction leads to the
creation of generations of O(6) kinks with masses close to $M_1 + M_2$  
and formation of two generations of the vector particles: one with
masses close to $2M_1$ and the other one with masses close to $2M_2$.  
\end{itemize}

We use the two stage RG. Let $M_3$ is the mass scale of the O(3) GN
model in the $-$ sector. At this energy the renormalization of $g_{-}$
coupling terminates and a nonzero average is formed: 
\bea
\ri\la\xi^b_R(M_3^{-1})\xi^b_L(0)\ra \approx -M_3  
\eea
Replacing $\bar\xi \xi$ operator by its average in the low energy effective action for $\chi$ fermions we obtain the effective action like (\ref{spin0})
with the UV cut-off $M_3$:
\bea
{\cal L}_{eff} = \frac{\ri}{2}\bar\chi_a\gamma_{\mu}\p_{\mu}\chi_a + \ri M_0\bar\chi_a\chi_a - g_{+}(M_3)(\bar\chi_a\chi_a)^2 \label{spin1}
\eea
where $M_0 = \ri g_X(M_3)\la\xi^b_R(M_3^{-1})\xi^b_L(0)\ra$. The form
of this Hamiltonian suggests that the vector triplet in the $-$ sector
is stabilized. For the procedure to be self-consistent the mass of
this vector particle should be  $M_{+} << M_3$. Let us assume that
$g_+(M_3)$ is still positive, which is the case at sufficiently small
$g_X(0)$. Then the downward  renormalization of $g_+$ will continue
till the energy $M_+$ (the resulting mass for $\chi$ fermions). The
latter   is determined by the equation  
\bea
M_{+} = \frac{M_0}{1 + \frac{g_+(M_3^{-1})}{4\pi}\ln(M_3/M_+)}.
\eea
% Setting $g_-(M_3^{-1}) \approx -1$ (strong coupling) we get 
%\bea
%g_+(M_3^{-1}) \approx (2C_{1/2} + |C_{-1/2}|)^{-2}
%\eea
We see that the presence of the  repulsion in the $+$ channel can
further reduce the mass of the vector particle. The model (\ref{spin1})
was introduced in Ref.~\onlinecite{spin1} to describe low-energy properties of
the spin S=1 Heisenberg antiferromagnetic chain.

\subsection{4 +2 splitting}

The trickiest and at the same time the most realistic case is when
the modes split as 2+4 with repulsive interaction between the four
magnetic modes. In this case neither group can produce a mass gap and
the gap is generated solely by the interaction. This happens, for
instance,  in the model with  no inter-chain tunneling in the presence
of the density-density interaction. Then only the $2k_F$ components
of the density interact leading  to the arrangement 
\[
g_{\s,+} = g_{\s,-} \equiv v > 0, g_{c,ss} = g_{c,st} \equiv g
\]
where $v$ originates from the intra-chain exchange interaction. The
O(6) symmetry is split as O(4)$\times$U(1) 
\bea
\ri g\cos\beta\Theta_c^{(-)}\sum_{a=0}^3\chi_R^a\chi_L^a +
v(\ri\chi_R^a\chi_L^a)^2. 
\eea
The RG equations
\bea
\dot g_{\rho,-} = - 4g^2, \dot v = -2v^2 - 2g^2, \dot g = -g(g_{\rho,-} + 3v)
\eea
where $\beta^2/4\pi = 1 + g_{\rho,-}/2\pi$. In the realistic case of
repulsion $g_{\rho,-} < 0$ the system scales to strong coupling with
$v$ changing sign. Assuming that $-g_{\rho,-}$ is the largest among
the coupling constants and neglecting its renormalization we arrive to
the approximate solution
\bea
g = g(0)\re^{|g_{\rho,-}|t}, ~~ v = v(0) -
\frac{g^2(0)}{|g_{\rho,-}|}(\re^{2|g_{\rho,-}|t} -1), 
\eea
such that $v$ changes sign at 
\bea
t^* \approx
\frac{1}{2|g_{\rho,-}|}\ln\left[\frac{|g_{\rho,-}|v(0)}{g^2(0)} + 1\right], 
\eea
which fixes the upper limit for the gaps ${E^*}^2/\Lambda \sim
\Lambda\re^{-2t^*}$. This corresponds to the scenario when the mass is
generated by the effective attraction among the fermions.  

At small $d =\beta^2/4\pi < 1/2$ another scenario can be realized. To
understand the spectrum we resort to $1/N$ approximation where $N$ is
the number of fermion species. Assuming that the fermion masses are
larger than the boson ones, we integrate over fermions and obtain the
following effective potential for  
$\Theta_c^{(-)} \equiv \phi$ field: 
\bea
V = -\frac{g^2N}{4\pi}\cos^2(\beta\phi)
\ln\left[\frac{\Lambda}{g|\cos(\beta\phi)|}\right]  
\eea
which can be approximated as a pure cosine. This potential is relevant
only when $d < 1/2$ and for $d > 1/4$ the spectrum contains only
kinks. Let us call their mass gap $m_{\phi}$. Then the  estimate for
the fermion gap is  
\bea
m_{\chi} \approx g\la\cos(\beta\phi)\ra = g
C_{\beta}\Lambda(m_{\phi}/\Lambda)^{d}  
\eea
where $C_{\beta}$ is a constant. The fermion gap is indeed
much larger than the kinks gap which justifies the integration over
the fermionic modes. Thus the spectrum in this case consists only of
the O(6) kinks. The kinks acquire quantum numbers from the spinor
representation of the O(6) group through attachment of the fermions
zero modes (see Section \ref{sec:semiclassical}).

\section{Correlation functions and experimental probes}
\label{sec:corrfunc}
 A detailed analysis of correlation functions is outside the scope
of this paper, so we restrict this discussion to some qualitative remarks.

 Among the available experimental methods the ones which probe
correlations functions at various frequencies and momenta are Angle
Resolved Photoemission (ARPES), inelastic neutron and X-ray
scattering.  All other probes measure  either local correlation
functions (Nuclear Magnetic Resonance (NMR), tunneling spectroscopy)
or correlation functions at zero momenta (optical conductivity, Raman
scattering).  

The only known system which effective Hamiltonian resembles the one
for a  single two-leg ladder are single wall carbon nanotubes. However,
though the doped  
nanotube is described by the effective Hamiltonian
(\ref{effectiveH}) \cite{Gogolin}, the values of the bare
couplings for the nanotubes of available size are such that the gaps
are extremely small. In other experimentally available systems, such
as the telephone number compound mentioned in Introduction, ladders
are packed into a three-dimensional (3D) arrangement. In that case
interactions of gapless charge modes from different ladders may lead
to 3D ordering.  

Let us discuss the behavior of correlation functions above 3D phase
transition (if such transition occurs). In this case both
$\Phi_c^{(+)}$ and $\Theta_c^{(+)}$ exponents have power law
correlation functions. Since these exponents enter the majority of
experimentally measurable  correlation functions, emission of all
massive excitations are accompanied by emission of gapless bosonic
modes. This makes the massive particles incoherent. The best one can
do at the circumstances is to look for operators whose correlation
functions contain matrix elements corresponding to emission of just
one massive particle (and, of course, a cascade of gapless bosons).  

Having this in mind, let us consider, for example, the SCd phase. In
this phase all $\s$ ($\mu$) fields are locked together with
$\cos[\sqrt\pi \Theta_c^{(-)}]$ ($\sin$). Then the operators with
matrix elements between the vacuum and a single-particle state of the
doublet vector particle are  the Fourier components of the particle
density at $2k_F^{(p)}$: 
\bea
&& \rho(2k_F^{(p)}) = \sum_{\s}R^+_{\s,p}L_{\s,p} \sim  \\
&& \ri\re^{\ri\sqrt\pi\Phi_c^{(+)}}\re^{\pm
  \ri\sqrt\pi\Phi_c^{(-)}}[\s_1\s_2\s_3\s_0 \mp \mu_1\mu_2\mu_3\mu_0]
\nonumber 
\eea
As far as the magnetic triplet is concerned, it is emitted at
$2(k_F^{(1)} + k_F^{(2)})$ by the magnetization operator 
\bea
&& S^a(2k_F) = R^+_{0,\s}\s^a_{\s\s'}L_{\pi,\s'} = \\
&& \frac{1}{2\pi a}\re^{\ri\sqrt\pi\Phi_c^{(+)}}
\re^{\ri\sqrt\pi\Theta_c^{(-)}}[N_a\mu_0 + M_a\s_0]\nonumber 
\eea
where 
\[
N^a = (\mu_1\mu_2\s_3, \mu_1\s_2\mu_3,\s_1\mu_2\mu_3), 
\]
\[
M^a = (\s_1\s_2\mu_3, \s_1\mu_2\s_3,\mu_1\s_2\s_3)
\]
The corresponding quasi-coherent peaks disappear below $T_c$ since
the $\Theta_c^{(+)}$ is locked and correlations of
$\re^{\ri\sqrt\pi\Phi_c^{(+)}}$ become short-ranged.  However, if the
SCd phase is replaced at stronger interactions by the $4k_F$ ordered
Wigner crystal (recall the discussion in Section II B), it is
$\Phi_c^{(+)}$ field which is locked. Then the all the above peaks
become sharp (this was in fact observed in the telephone number
compound \cite{Girsh}).

\section{Conclusions}

 In this paper we made an attempt to outline the picture of the excitation 
spectrum of a field theory with a symmetry of a doped two-leg ladder,
namely the anisotropic $O(6)$ Gross-Neveu model. This model is not
integrable and then it is not possible to obtain exact results except
for some specific points in parameter space. We combined information
coming from the RG analysis with semiclassical methods and form-factor
perturbation theory. It follows from our analysis  that throughout
most of the phase diagram  the spectrum consists of degenerate
quartets of kinks and antikinks and the multiplet of vector particles
split as 3 + 2 +1. This basic picture experiences corrections when one
moves through the phase diagram. Namely, in some areas of the phase
diagram  the splitting is extremely small, while in some others it
may become so large that some multiplets are
pushed in the continuum and become unstable. The phase diagram
presents different types of quantum critical points. At second order
transition lines masses of certain particles vanish. Very close to the
first order transition line additional generations of kinks
emerge. Strong interactions in some sectors may generate additional
bound states (like breathers in the asymmetric charge sector).  

As we have mentioned many times throughout the text, one potential
application of this theory is 'telephone number' compound
Sr$_{14-x}$Ca$_x$Cu$_{24}$O$_{41}$. This is undoubtedly a strongly
correlated system. The measurements of low-frequency dielectric and
optical response  demonstrate existence of a weakly pinned  phason
mode \cite{gorshunov},\cite{girsh} which we  identified as
$\Phi_c^{(+)}$ mode. At the same time such probes as NMR \cite{imai},
inelastic neutron scattering \cite{eccleston} and ARPES \cite{arpes}
show gaps in all other parts of the spectrum.  This is in agreement
with existing theoretical understanding of the ladder materials.

The question of validity of the  quasi-one-dimensional 
field theory description is decided by
(i) comparison between  the values of the gaps and  the bandwidth and
by (ii) presence of essentially one-dimensional effects, such as
different gap values for different channels. The field theory is valid
when the gaps are small compared to the bandwidth. 
The gaps extracted from the ARPES measurements were
obtained  only for $x =0$ where the number of holes is apparently
rather small (though not zero, since according to
Ref.~\onlinecite{girsh} the  gapless CDW mode exists at this
concentration). The ARPES  shows the single-electron gap $\sim 0.3$eV and
the  bandwidth $\sim 1.2$eV \cite{arpes}.  At the same time the
neutron scattering (also available only for $x =0$) gives the spin
triplet gap $\sim 32$ meV with the bandwidth for the spin excitations
$\sim 200$ meV. NMR measurements done in a broad range of 
$Ca$ concentrations show that the spin gap decreases with doping and
becomes less than 200K at $x > 3$. The absence of temperature
saturation of the magnetic susceptibility indicates a crossover to
the paramagnetic regime meaning that gaps for non-magnetic
excitations also become smaller at these $x$. Therefore the gap/bandwidth ratios are sufficiently small for the field theory description to be valid.

 On the other hand, the  optical
conductivity measurements indicate a presence of essentially one-dimensional effects. They show a strong peak at radio frequencies
(presumably coming from the phason mode) and a threshold at infrared
frequencies (the so-called CDW gap). If CDW state would form as a
result of Fermi surface  instability, as it happens in
three-dimensional systems with nested Fermi surfaces, this  gap would
be twice as large as the single-particle gap measured by ARPES. It
would also coincide with the spin gap. However, according to 
Ref.~\onlinecite{gorsh2}, the values of the CDW gaps at $x=0,3$ and 9 are
130 meV, 110 meV and 3 meV respectively which is either several times
larger or much smaller than the spin gap. From this analysis we conclude that 
 though detailed comparison between theory and experiment would be
premature,  the field theory
description of telephone number compound is a reasonable
approach ( as far as the system remains strongly one-dimensional which
probably corresponds to $x < 9$).

 We believe that the present experiments allow one to determine the part of the phase diagram where the telephone number compound is located. The recent X-ray 
measurements show that the holes may crystallize in a 
three-dimensional Wigner crystal \cite{abba2,abba3} (the use of the term is explained in \cite{wigner}). Since the
particle peaks sharpen below the transition, we take it as an
indication that this is a 4$k_F$ Wigner crystal replacing the SCd
phase, as described in Section \ref{sec:phases}. Such crystal exists only if the electron-electron interaction has a long range tail. This points towards the standard Coulomb interaction as the primary agent of its formation.  Such interaction may lead to relatively small  values of $\beta^2$
and consequently to new bound states. It is just possible, that one
such bound state (a breather in the $\Theta_c^{(-)}$ sector) appears
as a sharp peak in the Raman scattering experiments \cite{raman}.

%\begine{Acknowledgements}

\begin{acknowledgements}
We are  grateful to P. Azaria, 
M. J. Bhaseen, G. Blumberg, F. Essler, M. Fabrizio, E. Gava, T. Grava,
R. M. Konik, P. Lecheminant, 
S. Lukyanov,  G. Mussardo, A. A. Nersesyan,  T. M. Rice,
G. Sotkov, F. A. Smirnov  and A. B. Zamolodchikov with
whom various aspects of this work were discussed on different stages,
for help and interest to the work.  AMT  acknowledges the support
from US DOE under contract number DE-AC02 -98 CH 10886 and is grateful to
Abdus Salam ICTP and Princeton University for hospitality. DC
acknowledges hospitality and  support from Institute for Strongly
Correlated and Complex Systems at BNL. 
\end{acknowledgements}

\appendix
\section{Bosonization}

We adopt the following notations:
\bea
&& R_{p,\s} = 
\frac{\eta_{p\s}}{\sqrt{2\pi a_0}}\re^{-\ri\sqrt{4\pi}\varphi_{p\s}},
\nonumber\\
&& L_{p,\s} = 
\frac{\eta_{p\s}}{\sqrt{2\pi a_0}}\re^{\ri\sqrt{4\pi}\bar\varphi_{p\s}},
\eea
where $p = \pm 1$ and $\varphi$ and $\bar\varphi$ are bosonic fields
with right and left chirality. The Klein factors satisfy
anticommutation relations of the O(4) Clifford algebra 
\[
\{\eta_a,\eta_b\} = \delta_{ab}
\]
and can be chosen as follows \cite{marston}:
\bea
\eta_{-1\s}\eta_{1\s} = \ri, ~~ \eta_{p\uparrow}\eta_{p\downarrow} 
=\ri(-1)^{(p +1)/2}.
\eea
The chiral bosonic fields are decomposed into the normal modes as follows:
\bea
\varphi_{p\s} = \frac{1}{2}\left\{[\phi_c^{(+)} + p\phi_c^{(-)}] +
\s[\phi_s^{(+)} + p\phi_s^{(-)}]\right\}
\eea
with the same decomposition for $\bar\varphi$. The bosonic field,
$\Phi$, and its dual, $\Theta$ as usual are
\be
\Phi_{s,c}^{(\pm)}= \phi^{(\pm)}_{s,c}+\bar\phi^{(\pm)}_{s,c}\,,\,\,\, 
\Theta_a^{(\pm)}= \phi^{(\pm)}_{s,c}-\bar\phi^{(\pm)}_{s,c}\,.
\ee

The Ising model order and disorder parameter fields $\s,\mu$ are related to the bosonic fields as follows:
\bea
&& \cos[\sqrt{\pi}\Phi_s^{(+)}] = \s_1\s_2, ~~
\sin[\sqrt{\pi}\Phi_s^{(+)}] = \mu_1\mu_2, \nonumber\\ 
&& \cos[\sqrt{\pi}\Theta_s^{(+)}] = \mu_1\s_2, ~~
\sin[\sqrt{\pi}\Theta_s^{(+)}] = \s_1\mu_2,  
\eea
with the similar formulas relating $\Phi_s^{(-)}, \Theta_s^{(-)}$ to
$\s_3,\s_0, \mu_3,\mu_0$.

\section{Semiclassical analysis of generalized $O(2n)$ GN models}
\label{app:semiclassical}
In order to have a more intuitive picture of the effect of perturbations 
(\ref{pert1}) and (\ref{pert2}) on GN models we repeat the semiclassical
analysis of Sec.~\ref{sec:semiclassical} for $N$ and $M$ even, $N=2n$ and 
$M=2m$,
following Ref.~\onlinecite{shankar-witten}.
The advantage of this assumption lies in the fact the Majorana fermions can be
bosonized in couples
and then the potential of the $O(2n)$ GN model (\ref{GN}) takes the form 
\be 
V=-g' \left (\sum_{i=1}^n \cos(\sqrt{4\pi} \phi_i) \right )^2.
\label{bosonized1}
\ee
It has two families of minima, $\phi_i=\sqrt{\pi} n$ and 
$\phi_i=\sqrt{\pi} (n+1/2)$ that are usually referred to as positive and
negative vacua because they correspond to positive and negative values of
$\sum_a \psi_R^a \psi_L^a= \sum_i \cos \sqrt{4 \pi} \phi_i$. Fermionic 
excitations (when stable) correspond to configurations 
interpolating between minima that belong to the same family, for
instance from a configuration $(0,\ldots,0)$ at $x \to -\infty$ to  $(\pm
\sqrt{\pi},0,\ldots,0)$, at $x \to +\infty$. On the other hand, kinks
interpolate between minima of different families, for example, from 
$(0,\ldots,0)$ for $x\to -\infty$ to $(\pm \sqrt{\pi}/2,\ldots,\pm
\sqrt{\pi}/2)$ for $x\to +\infty$. It is easy to see that, while there are $N$
possible fermionic states, the number of kinks states is $2^n$. Please
note that since kinks interpolate  between positive and
negative vacua the mean value of $\psi_R^a \psi_L^a$ changes sign along a kink
configuration.  

From this picture it is clear that fermions can be considered as bound states
of kinks. If fact, for instance, an elementary excitation associated to the
transition $(0,\ldots,0) \to (\sqrt{\pi},0,\ldots,0)$ can be obtained also
with two transitions $(0,\ldots,0) \to (+\sqrt{\pi}/2,\ldots,+
\sqrt{\pi}/2) \to (\sqrt{\pi},0,\ldots,0)$, where each of the two jumps
corresponds to a kink. The stability of fermions against the decay into a pair
of kinks depends on $N$. 

Let us now first consider the effect of a simple perturbation of the form
\be 
V'=- \sum_i \cos \sqrt{4\pi} \phi_i.
\ee
Here the positive and negative minima are split and only the former remain
absolute minima. This implies that kinks
interpolating between different families are not present
anymore and the model has only fermionic excitations. This is a simple example
of confinement. 
If we now consider the model (\ref{pert2}), the effect of the perturbation on
the excitations of the two GN models, originally decoupled, is
somewhat similar. When the two models are decoupled each of them  
has the same families of minima described above. For instance, if we
introduce $\phi_i$ and $\sigma_i$ associated to the $\chi^a$s and $\xi^b$s
respectively, configurations like
$(\phi_1,\ldots,\phi_n;\sigma_1,\ldots,\sigma_n)=
(0,\ldots,0;\pm\sqrt{\pi}/2,\ldots,\pm\sqrt{\pi}/2)$ 
are absolute minima. Nevertheless this does not remain true in
presence of a perturbation of the form
\be 
\delta \tilde g_{a,b}  \cos \sqrt{4 \pi}\phi_a  \cos\sqrt{4\pi}\sigma_b
\ee
(here we assume that also the perturbation is such that it can be
written in a simple bosonic form). 
Again the kinks of the two decoupled GN models confine
and disappear from the spectrum. At the same time, together with
fermions, new kinks with the quantum
number of $O(N+M)$ GN appear. They interpolates between configurations like
$(0,\ldots,0;0,\ldots,0) \to (\pm\sqrt{\pi}/2,\ldots,\pm\sqrt{\pi}/2;
\pm\sqrt{\pi}/2,\ldots,\pm\sqrt{\pi}/2)$, that remain degenerate
minima also of the perturbed theory. Fermionic excitations of the two
models remain stable. 

Following the same arguments one can easily see that in
presence of perturbations of the form (\ref{pert1}) the $O(N)$ kinks remain
stable. In fact, the potential 
\be
\delta g_{a,b}  \cos \sqrt{4 \pi}\phi_a  \cos\sqrt{4\pi}\phi_b
\ee
does not lift the degeneracy of the minima of the unperturbed GN model.

\section{The exact solution of (5+1) and (3+1) models} 
\label{app:exactsol}

In this appendix we describe the S-matrix of two models, that we call
(5+1) and (3+1)-model, with $O(5)\times \mathbb{Z}_2$ and $O(3)\times
\mathbb{Z}_2$ symmetry respectively. The latter was introduced and
studied in detail in Ref.s~\onlinecite{Ts87,Andrei}. The central
charge in the UV computed with the Thermodynamics Bethe Ansatz is
$c_{5+1}=3$ and $c_{3+1}=2$. The relationship with the models of
interest in the paper is discussed in Sec.~\ref{sec:int.points}.
 
The spectrum of the (5+1)-model
consists of a singlet Majorana fermion with mass $m_0$,
a quintet of Majorana fermions with masses $\sqrt 3 M$ and a quartet
of kinks  $\equiv$ antikinks (spinor particles) with mass $M$. The
S-matrix for the (5+1) model is 
\bea
S_{5+1} = \left(
\begin{array}{ccc}
S_{vv} & S_{vs} & -I\\
S_{vs} & S_{ss} & \xi \delta_{\alpha}^{\bar\alpha} \\
-I & \xi \delta_{\alpha}^{\bar\alpha} & - 1
\end{array}
\right)
\eea
where 
\[
\xi(\theta) = \frac{\re^{3\theta} - \ri}{\re^{3\theta} + \ri}
\]
and $S^{vv}$, $S^{vs}$ and $S^{ss}$ are the O(5) S-matrices of vector
and spinor particles.  The above   $S$ matrices were found in
Ref.~\onlinecite{Ann} 
(for the vector particles) and \onlinecite{OgReshW} (for kinks). 
The spinor S-matrix has the following form:
\bea
S^{ss}(\theta) = f(\theta)\left[\hat P_{asym} + \frac{\theta +
    \ri\pi/3}{\theta - \ri\pi/3}\hat P_v + \frac{\theta +
    \ri\pi}{\theta - \ri\pi}\hat P_0\right], \label{spinor} 
\eea
where $P_0, P_v, P_{asym}$ represent projectors onto singlet, vector
and antisymmetric tensor representations and  
\bea
f(\theta) = \frac{\Gamma\left(1 +
  \frac{\ri\theta}{2\pi}\right)}{\Gamma\left(1 -
  \frac{\ri\theta}{2\pi}\right)} 
\frac{\Gamma\left(\frac{1}{2} -
  \frac{\ri\theta}{2\pi}\right)}{\Gamma\left(\frac{1}{2} +
  \frac{\ri\theta}{2\pi}\right)}\frac{\Gamma\left(\frac{5}{6} -
  \frac{\ri\theta}{2\pi}\right)}{\Gamma\left(\frac{5}{6} +
  \frac{\ri\theta}{2\pi}\right)} 
\frac{\Gamma\left(\frac{1}{3} + 
  \frac{\ri\theta}{2\pi}\right)}{\Gamma\left(\frac{1}{3} -
  \frac{\ri\theta}{2\pi}\right)}. \nonumber 
\eea
The  S-matrix has one  physical pole on the physical strip ($0 < \Im
m\theta < \pi$)  at $\theta = \ri\pi/3$ corresponding to the vector
particle and one unphysical at $2\pi\ri/3$. 
Each of the S-matrices is crossing symmetric, including the scalar
factor $\xi$: $\xi(\theta) = - \xi(\ri\pi - \theta)$.  

For the (3+1) model we have a similar structure:
\bea
S_{3+1} = \left(
\begin{array}{cc}
 [S^{SU(2)}]_{\alpha,\beta}^{\bar\alpha,\bar\beta} & \xi
 \delta_{\alpha}^{\bar\alpha} \\ 
\xi \delta_{\alpha}^{\bar\alpha} & - 1
\end{array}
\right)
\eea
where 
\[
\xi(\theta) = \frac{\re^{\theta} - \ri}{\re^{\theta} + \ri}
\]
and $S^{SU(2)}$ is the S-matrix of SU(2) Thirring model solitons.

\section{Solution of the linearized RG equations}
\label{appendix:lin}
The RG equations (\ref{RG}) can be simplified for $g_a=g+\delta g_a$
with $g/\delta g_a \ll 1$. Defining for simplicity of notations
$\delta g_a \equiv x_a$ we find that, to first order in ${\bf x}$
Eq.s~(\ref{RG}) take the form 
\bea 
\label{diff.x}
\frac{\dot {\bf x}}{2g} = - D {\bf x} 
\eea 
where
\bea D = \left(
\begin{array}{ccccc}
0 & 1 & 0 & 3 & 0\\ 1/2&1/2& 3/2&3/2&0\\ 0 &1 & 1 & 1& 1\\ 1/2 &1/2
&1/2 &3/2 & 1 
\\ 0 & 0 & 1 &2 & 1
\end{array}
\right)
\label{D}
\eea 
and $g$ is solution of $\dot g=-4g^2$ and has the form
\be
g(t)=\frac{1}{4 t -g_o^{-1}}.
\ee
It is convenient to introduce ${\bf y}$ 
\be
{\bf x}=T {\bf y},
\label{x.y}
\ee
that satisfies the equation
\be
 \dot{\bf y} = -T^{-1}D T ~{\bf y} ~2g(t),~~~{\bf y}(0)=T^{-1} {\bf
   x}(0)\equiv A_a , 
\label{diff.y}
\ee
where $T$ 
\be
T=\left (
\begin{array}{ccccc}
-3 & -6 & -1 & 2 & 1\\ 3& 3 & -1 & -1 &1 \\ -2 &-2 & 0  & -2& 1\\ 0 &1
& 0 &1  & 1
\\ 1 & 0 & 1 &0 & 1
\end{array} \right )
\ee
is chosen such that $T^{-1}D T={\rm diag}
(\lambda_1,\lambda_2,\lambda_3,\lambda_4,\lambda_5)$,
with $\lambda$s being the eigenvalues of (\ref{D}),
$\lambda_1=\lambda_2=-1$, $\lambda_3=\lambda_4=1$, $\lambda_5=4$. 
Since the eigenvectors of degenerate eigenvalues are linearly
independent all solutions of (\ref{diff.y}) have the form 
\be
y_a =A_a \exp [-\lambda_a l(t)]
\ee
where 
\be
l(t)=\frac{1}{2}\log(g_0^{-1}) -\frac{1}{2}
\log(g_0^{-1} -4 t).
\ee
From this it follows that
\bea
&&y_a =
%A_a\left ( \frac{4 t^*}{4(t^*-t)} \right )^{-\lambda_a/2}\\
A_a \left (\frac{1}{g_0} \right )^{\lambda_a/2}
\left(\log (\varepsilon/m)+1  \right) ^{-\lambda_a/2},
\eea
where the relationship between $t$ and $\varepsilon$ was used.
The solution of (\ref{diff.x}) can be obtained inverting (\ref{x.y}).


\begin{thebibliography}{99}
\bibitem{wigner} The term Wigner crystal is used to denote  an ordered state where the  spacial  order originates from the electron-electron interactions. In quasi-one-dimensional systems such ordering does not necessarily require that the average Coulomb energy is much greater than the kinetic one. Therefore the   density oscillations in quasi-1D Wigner crystals may be  almost sinusoidal, unlike in an isotropic case.
\end{thebibliography}
\end{document}